\documentclass[twocolumn]{aastex631}
\usepackage{natbib}
\usepackage{amsmath}
\usepackage{hyperref}

\newcommand{\nicer}{\textit{NICER}}

\newcommand{\xmm}{{\it XMM}}

\begin{document}
% \linenumbers

\title{Testing EMRI models for Quasi-Periodic Eruptions with \textcolor{black}{3.5 years of monitoring} eRO-QPE1}

\author[0000-0002-0568-6000]{Joheen Chakraborty}\thanks{joheen@mit.edu}
\affiliation{Department of Physics \& Kavli Institute for Astrophysics and Space Research, Massachusetts Institute of Technology, Cambridge, MA 02139, USA}

\author[0000-0003-4054-7978]{Riccardo Arcodia}\thanks{NASA Einstein Fellow}
\affiliation{Department of Physics \& Kavli Institute for Astrophysics and Space Research, Massachusetts Institute of Technology, Cambridge, MA 02139, USA}

\author[0000-0003-0172-0854]{Erin Kara}
\affiliation{Department of Physics \& Kavli Institute for Astrophysics and Space Research, Massachusetts Institute of Technology, Cambridge, MA 02139, USA}

\author[0000-0003-0707-4531]{Giovanni Miniutti}
\affiliation{Centro de Astrobiolog\'ia (CAB), CSIC-INTA, Camino Bajo del Castillo s/n, 28692 Villanueva de la Ca\~nada, Madrid, Spain}

\author[0000-0002-1329-658X]{Margherita Giustini}
\affiliation{Centro de Astrobiolog\'ia (CAB), CSIC-INTA, Camino Bajo del Castillo s/n, 28692 Villanueva de la Ca\~nada, Madrid, Spain}

\author[0000-0003-3906-4354]{Alexandra J. Tetarenko}\thanks{Former NASA Einstein Fellow}
\affiliation{Department of Physics and Astronomy, University of Lethbridge, Lethbridge, Alberta, T1K 3M4, Canada}
\affiliation{Department of Physics and Astronomy, Texas Tech University, Lubbock, TX 79409-1051, USA}

\author[0000-0003-2705-4941]{Lauren Rhodes}
\affiliation{Astrophysics, Department of Physics, University of Oxford, Denys Wilkinson Building, Keble Road, Oxford, OX1 3RH, UK} 

\author[0000-0002-8400-0969]{Alessia Franchini}
\affiliation{Center for Theoretical Astrophysics and Cosmology, Institute for Computational Science, University of Zurich,
Winterthurerstrasse 190, CH-8057 Zürich, Switzerland}

\author[0000-0001-7889-6810]{Matteo Bonetti}
\affiliation{Dipartimento di Fisica ``G. Occhialini'', Universit\`a degli Studi di Milano-Bicocca, Piazza della Scienza 3, I-20126 Milano, Italy}

\author[0000-0002-7226-836X]{Kevin B. Burdge}
\affiliation{Department of Physics \& Kavli Institute for Astrophysics and Space Research, Massachusetts Institute of Technology, Cambridge, MA 02139, USA}

\author[0000-0003-3441-8299]{Adelle J. Goodwin}
\affiliation{International Centre for Radio Astronomy Research -- Curtin University, GPO Box U1987, Perth, WA 6845, Australia}

\author{Thomas J. Maccarone}
\affiliation{Department of Physics and Astronomy, Texas Tech University, Lubbock, TX 79409-1051, USA}

\author[0000-0002-0761-0130]{Andrea Merloni}
\affiliation{Max-Planck-Institut f{\"u}r extraterrestrische Physik, Gießenbachstra{\ss}e 1, D-85748 Garching bei M\"unchen , Germany}

\author[0000-0003-0293-3608]{Gabriele Ponti}
\affiliation{INAF-Osservatorio Astronomico di Brera, Via E. Bianchi 46, I-23807 Merate (LC), Italy}
\affiliation{Max-Planck-Institut f{\"u}r extraterrestrische Physik, Gießenbachstra{\ss}e 1, D-85748 Garching bei M\"unchen , Germany}

\author[0000-0003-4815-0481]{Ronald A. Remillard}
\affiliation{Department of Physics \& Kavli Institute for Astrophysics and Space Research, Massachusetts Institute of Technology, Cambridge, MA 02139, USA}

\author[0000-0002-4912-2477]{Richard D. Saxton}
\affiliation{Telespazio UK for ESA, ESAC, Camino Bajo del Castillo s/n, 28692 Villanueva de la Ca\~nada, Madrid, Spain}

\begin{abstract}
Quasi-Periodic Eruptions (QPEs) are luminous X-ray outbursts recurring on hour timescales, observed from the nuclei of a growing handful of nearby low-mass galaxies. Their physical origin is still debated, and usually modeled as (a) accretion disk instabilities or (b) interaction of a supermassive black hole (SMBH) with a lower mass companion in an extreme mass-ratio inspiral (EMRI). EMRI models can be tested with several predictions related to the short- and long-term behavior of QPEs. In this study, we report on the ongoing \textcolor{black}{3.5}-year \nicer\ \textcolor{black}{and \textit{XMM-Newton}} monitoring campaign of eRO-QPE1, which is known to exhibit erratic QPEs that have been challenging for the simplest EMRI models to explain. We report 1) complex, non-monotonic evolution in the long-term trends of QPE energy output and inferred emitting area; 2) the disappearance of the QPEs (within \nicer\ detectability) in October 2023, then reappearance \textcolor{black}{by January 2024 at a luminosity $\sim$100x fainter (and temperature $\sim$3x cooler) than initial discovery}; 3) radio non-detections with MeerKAT and VLA observations partly contemporaneous with our \nicer\ campaign (though not during outbursts); and 4) the presence of a possible $\sim$6-day modulation of the QPE timing residuals, which aligns with the expected nodal precession timescale of the underlying accretion disk. Our results tentatively support EMRI-disk collision models powering the QPEs, and we demonstrate that the timing modulation of QPEs may be used to jointly constrain the SMBH spin and disk density profile. 
\end{abstract}


\section{Introduction} \label{sec:intro}
Quasi-Periodic Eruptions (QPEs) are a rare class of extragalactic transient characterized by high-amplitude, repeating soft X-ray flares with recurrence times of hours. They were first discovered in the nucleus of GSN 069 \citep{Miniutti19}, and since then there have been \textcolor{black}{five} more confirmed QPE hosts \citep{Giustini20,Arcodia21,Arcodia24} and two candidates \citep{Chakraborty21,Quintin23}. Related properties, including recurring large X-ray outbursts and super-soft thermal spectra, are also observed from a handful of other \textcolor{black}{supermassive black holes (SMBHs)} \citep{Terashima12,Tiengo22,Evans23,Guolo23}.

QPEs have been observed so far from the nuclei of nearby low-mass galaxies ($M_*\approx10^{9-9.5}\,$M$_{\odot}$, \citet{Arcodia21}), hosting black holes of $M_{BH}\approx10^{5-6.6}\,$M$_{\odot}$ \citep{Wevers22}. Contrary to other kinds of X-ray transients which show multi-wavelength counterparts, the flares from QPEs have so far been observed only in the soft X-rays (though it is possible that other wavelengths are dominated by the host galaxy or the disk itself). Their spectra are consistent with the exponentially decaying Wien tail of a blackbody ($kT\approx100-200\,$eV), with soft X-ray luminosities reaching up to $10^{42}-10^{43}$\,erg\,s$^{-1}$ ($\sim 0.01-0.1 L_{Edd}$). In eruption, these sources increase their count rate by up to 10-150 times compared to the quiescence state in between flares. Quiescence, when detected, is ultra-soft ($kT\approx50-80\,$eV) and it is interpreted as the emission from the accretion disk around the central massive black hole. 

Currently, the origin of QPEs is still debated. Existing models separate broadly into (a) recurring limit-cycle instabilities within the SMBH accretion disk \citep{Raj21,Pan22,Pan23,Kaur23,Sniegowska23}; or (b) the interaction of the SMBH with a lower-mass orbiting companion, which allows many different prescriptions for precisely how the X-ray emission is produced \citep{Xian21,Sukova21,King22,Metzger22,Krolik22,Zhao22,Linial23a,Franchini23,Linial23b,Tagawa23,linial24}. The latter class of models has generated much excitement, as they may reveal QPEs to be the first-observed electromagnetic counterparts of extreme mass-ratio inspirals (EMRIs) \citep{Arcodia21}, a source of millihertz gravitational waves \textcolor{black}{in the sensitivity band of} upcoming space-based gravitational wave detectors (though the detectability of the thus-known population is limited, \citet{Chen22}).

Finally, multiple lines of evidence have linked QPEs to Tidal Disruption Events (TDEs). The QPEs in GSN 069 \textcolor{black}{and eRO-QPE3 are seen alongside} a long-term decay of the quiescent flux level \citep{Miniutti19,Arcodia24}. Two QPE candidates were also discovered in TDEs, which were initially detected by \xmm\ Slew \citep{Chakraborty21} and the Zwicky Transient Facility \citep{Quintin23}. GSN 069 shows an abnormal C/N ratio \citep{Sheng21} and a compact nuclear [O III] region suggesting a young accretion system \citep{Patra23}. Although optical spectra of QPE hosts show signatures for nuclear ionization, no QPE host nuclei are associated with broad optical/UV emission lines, which may suggest the disks around their SMBHs are too compact to support a mature \textcolor{black}{broad line region} \citep{Wevers22}. Recent modelling efforts have begun to explicitly account for the coincidence with TDEs to explain the QPE phenomenon \citep{Linial23b,Franchini23}, as the rarity of TDEs themselves is an extremely strong constraint of the prerequisite conditions needed to produce QPEs. Continued monitoring of known sources \citep{Miniutti23a} and late-time follow-up of TDE candidates to identify unusual X-ray variability will be a powerful tool in determining the relationship between TDEs and QPEs.

With the initial sample of QPE sources now growing to multi-year observational baselines, we are able to begin studying the secular evolution of these sources, which allows important tests of theoretical models. \citet{Miniutti23a} and \citet{Miniutti23b} studied the 12-year \xmm\ dataset of GSN 069 (with QPEs appearing in the past 5 years), finding significant long-term variability in the quiescence luminosity. \textcolor{black}{They also found coupling of the QPE properties to the quiescent accretion disk: as accretion rate increases, the QPE temperature and amplitude both decrease, with the QPE temperature asymptotically approaching the disk temperature}. Giustini et al. (in prep) found modest variations of the quiescent luminosity of RX J1301.9+2747, and the constant presence of QPEs with a complex pattern of variability, over more than 20 years of X-ray observations.

Here, we report on the \textcolor{black}{3.5}-year \nicer\ \textcolor{black}{\textit{and XMM-Newton}} dataset of eRO-QPE1, a source initially reported in \citet{Arcodia21}. The strikingly complex flare patterns and unpredictable recurrence times were immediately noteworthy, and the comparatively erratic QPE behavior has presented a challenge for the simplest EMRI-based models which predict roughly regular outbursts on the orbital timescale. The initial \xmm\ observations of this source were studied in-depth by \citet{Arcodia22}, who found evidence for overlapping QPEs and a similar $L-kT$ hysteresis pattern as seen in GSN 069. Our long-baseline \nicer\ dataset allows us to probe secular evolution of the QPEs and their timing properties across three years and a total of 92 flares. In Section~\ref{sec:methods} we report \textcolor{black}{on our analysis procedures, including a non-standard \nicer\ data reduction process}, and subsequent analysis methods. In Section~\ref{sec:results} we report the results of time-resolved spectroscopy and timing, as well as implications for theoretical models.

\section{Methods}
\label{sec:methods}

\subsection{\nicer\ light curves and spectra}
\label{subsec:nicer}

\begin{figure*}
    \centering
    \includegraphics[width=0.95\linewidth]{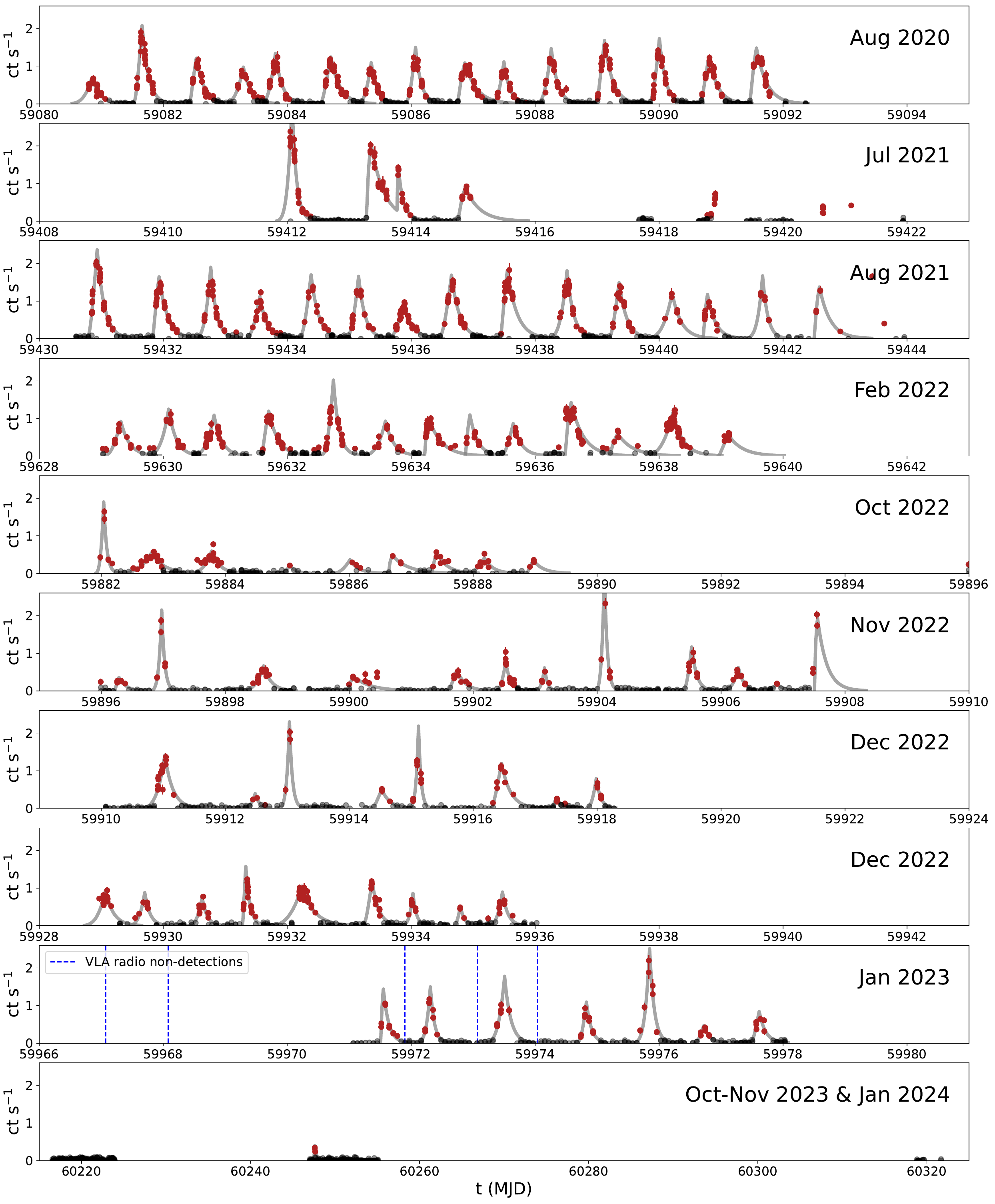}
    \caption{\nicer\ background-subtracted light curves of eRO-QPE1 spanning three years, derived by fitting the source with a blackbody spectral model (\texttt{tbabs}$\times$\texttt{zbbody}) in each 200-second GTI. \textcolor{black}{Red} points denote source detections, while \textcolor{black}{black} points indicate GTIs in which a blackbody component was not required at $>1\sigma$. There are 92 flares, which we fit with the exponential model described in Sec. 2.2 (gray lines).}
    \label{fig:lc}
\end{figure*}

eRO-QPE1 was observed in \textcolor{black}{141} \nicer\ observations (PI: Arcodia) for a total of 964.2 kiloseconds between August 2020-\textcolor{black}{January 2024}. The data were processed using \texttt{HEAsoft} v6.32.1 and \texttt{NICERDAS} v11a. The standard procedure of using \texttt{nicerl3} to generate source light curves was insufficient here, due to the low signal/noise ratio and variable background; thus we devised a non-standard procedure, detailed here.

Reliably estimating light curves for faint sources like eRO-QPE1, in which the source count rate is generally comparable to (or less than) the background, presents a challenge for \nicer. In most observations, the combination of background counts produced by optical loading (at low energies) and particle interactions with the Si detectors (at high energies) produces a higher count rate than the source itself. This background level can also vary slightly within observations, creating a degeneracy of the observed total variability between source and background components---because \nicer\ is a non-imaging instrument, variability cannot be resolved directly into source and background components, instead requiring other metrics to estimate total background contribution. We approach the problem using time-resolved spectroscopy by splitting the data into many Good Time Intervals (GTIs), then spectroscopically estimating the source and background components separately in each GTI.

\begin{figure*}
    \centering
    \includegraphics[width=0.95\linewidth]{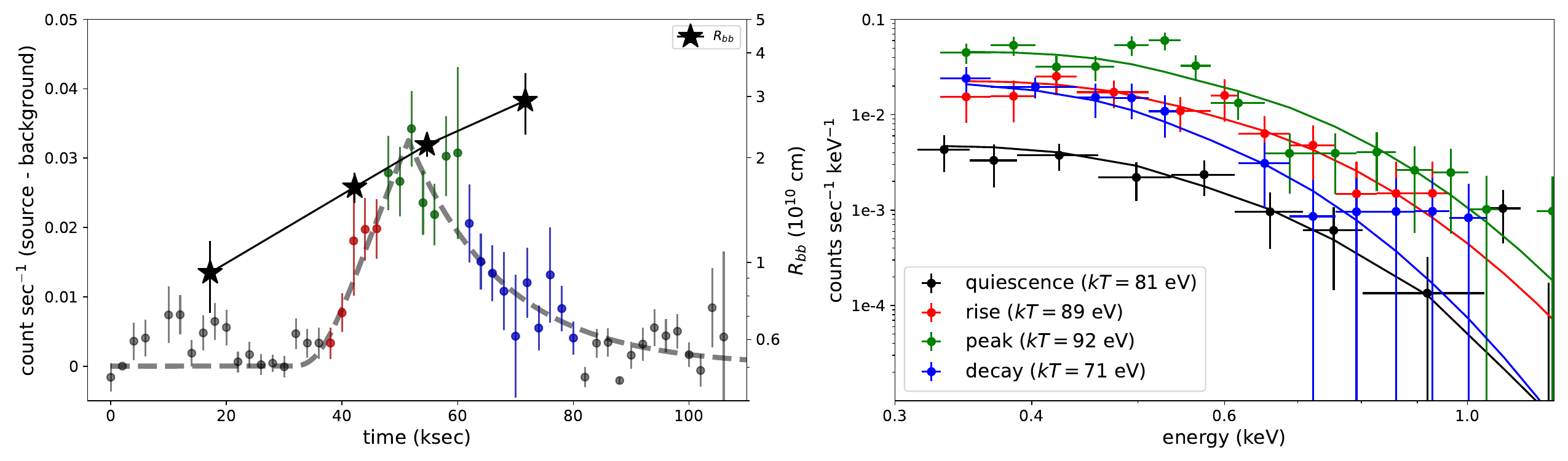}
    \caption{\textcolor{black}{\textbf{Left:} Light curve showing the QPE observed on Jan 5, 2024 with \textit{XMM-Newton}, with the quiescence/rise/peak/decay phases denoted by black/red/green/blue points. Overplotted is the blackbody emitting radius (see Sec.~\ref{subsec:evolution_results}). \textbf{Right:} phase-resolved spectra showing quiescence, rise, peak, and decay, along with \texttt{tbabs}$\times$\texttt{zbbody} blackbody fits.}}
    \label{fig:xmm}
\end{figure*}

First we use \texttt{nimaketime}, with custom filtering choices of unrestricted undershoot (\texttt{underonly\_range}=*-*) and overshoot rates (\texttt{overonly\_range}=*-*), as well as per-FPM and per-MPU autoscreening disabled to prevent unnecessarily aggressive event filtering. We then split the intervals produced by \texttt{nimaketime} into GTIs of maximum 200 seconds to allow time-resolved estimation of the variable background. In each GTI, we manually discard focal plane modules (FPMs) with 0-0.2 or 5-15 keV count rates $>4\sigma$ higher than the average across all GTIs within the OBSID, or above an absolute threshold of 20 counts sec$^{-1}$. These cutoffs were chosen because eRO-QPE1 is super-soft ($kT_{max} \sim 120$ eV) and faint ($\sim$2 counts sec$^{-1}$ at peak). Thus the 5+ keV band is entirely background-dominated (so 20+ counts sec$^{-1}$ signifies extremely high particle-induced background), whereas the 0-0.2 keV band is undershoot-dominated (so 20+ counts sec$^{-1}$ signifies extremely severe light leak conditions).

After screening the event lists, we used the \texttt{SCORPEON}\footnote{\href{https://heasarc.gsfc.nasa.gov/lheasoft/ftools/headas/niscorpeon.html}{\texttt{https://heasarc.gsfc.nasa.gov/lheasoft/ftools/\\headas/niscorpeon.html}}} template-based background model to estimate the contribution from astrophysical and non X-ray backgrounds separately for each GTI. We fit the entire broadband (0.2-15 keV) array counts with the PyXspec\footnote{\href{https://heasarc.gsfc.nasa.gov/docs/xanadu/xspec/python/html/index.html}{\texttt{https://heasarc.gsfc.nasa.gov/docs/xanadu/xspec/\\python/html/index.html}}} interface to \texttt{XSPEC} \citep{Arnaud96}. We leave the Solar Wind Charge Exchange (SWCX) oxygen emission line normalizations free to vary to account for partially ionized oxygen fluorescence from the solar wind (which is particularly important during \nicer\ dayside GTIs). Along with the \texttt{SCORPEON} background we fit each GTI with a source model represented by \texttt{tbabs}$\times$\texttt{zbbody} (motivated by the pure thermal-like emission spectrum of the source, \citet{Arcodia21}) \textcolor{black}{with $z=0.0505$ fixed to the host galaxy redshift and $N_H=2.23\times 10^{20}$ cm$^{-2}$ fixed to the line-of-sight galactic column density}, resulting in a distribution of broadband fit statistics (cstat/dof) with $\mu = 1.2$, $\sigma=0.3$. We sum only the counts contained within the \texttt{zbbody} model to create the background-subtracted light curves presented in Fig.~\ref{fig:lc}. We consider a source ``detection" as any GTI in which the blackbody normalization is $>$1$\sigma$ inconsistent with zero, i.e. a non-background component is required by the fit at the $1\sigma$ level. See Appendix~\ref{appendix:detection} for examples of robust and marginal detections, as well as further discussion of the source and background spectral components.

We note that \nicer\ was not able to detect the quiescent disk emission in eRO-QPE1 at any epoch. As eRO-QPE1 is the most distant known QPE ($z=0.0505$), the quiescence is detected only at $1\sigma$ by \xmm\ \citep{Arcodia21} despite a similar $L_X$ to other QPE hosts.

We also perform time-resolved spectroscopy on relative-intensity bins to track the evolution of the QPE bolometric luminosity ($L_{bol}$), temperature  ($kT$), and the inferred blackbody radius ($R_{bb}$) over the different phases of each flare (Fig.~\ref{fig:rbb_kT}). We followed the same filtering and background estimation procedures described above, then grouped all GTIs with a source detection into 5 relative-intensity bins for each QPE, based on its flux compared to peak: 1-80\% $F_{peak}$, 80-100\%, 100-80\%, 80-50\%, and 50-1\%. We choose 2 rise bins and 3 decay relative-intensity bins, because the faster rise means they are typically more poorly sampled. Not all QPEs are well-sampled across all 5 segments, because \nicer\ did not always fully sample the rise and decay (Figs.~\ref{fig:rbb_kT},~\ref{fig:rbb_all}).

\subsection{XMM-Newton}
\label{subsec:xmm}
\textcolor{black}{
Upon observing the possible re-emergence of QPEs in Nov 2023 with \textit{NICER} (Fig.~\ref{fig:lc}), we requested a 124\,ks Director's Discretionary Time (DDT) observation with \textit{XMM-Newton}, which was taken on Jan 5-6, 2024. The data were reduced with standard tools and prescriptions (XMM SAS v20.0.0 and HEAsoft v6.29). For eRO-QPE1, source products were extracted within a circle of 13.8" to avoid a nearby contaminant, while background was extracted from a source-free area within a circle of 39". We restrict our analysis to the EPIC-PN instrument due to its higher count rate compared to the MOS CCDs. Source and background rates were then extracted, and grouped into 2000-second bins to produce the background-subtracted light curve in Fig.~\ref{fig:xmm}. Phase-resolved spectra were extracted with \texttt{evselect}, represented by the differently-colored regions in Fig.~\ref{fig:xmm} (quiescence comprises the beginning and ending segments of the observation). We analyzed the spectra using \texttt{XSPEC} v12.13.1, fitting with the Cash statistic and using data from 0.2-1.5 keV (as the source is background-dominated at higher energies). We fit spectra using the same model as the NICER data (\texttt{tbabs}$\times$\texttt{zbbody}) discussed in Sec.~\ref{subsec:nicer}, with fitting results and comparison to previous \textit{XMM-Newton} results \citep{Arcodia21} reported in Table~\ref{tab:xmm}.}

\begin{table}
\begin{center}
\textcolor{black}{
\textit{XMM} Jan 2024
\begin{tabular}{c|c|c|c|c}
\hline
Phase & $kT$ & $L_{bol}$ &  cstat/dof & $R_{bb}$\\
 & (eV) & (erg s$^{-1}$) &  & ($10^{10}$ cm) \\
\toprule
Quiesc.  & $82^{+17}_{-14}$ & $5.1^{+1.5}_{-1.6}\times 10^{40}$    & 18/21 & $0.93^{+0.14}_{-0.14}$  \\
Rise        & $89^{+14}_{-11}$ & $2.2^{+0.04}_{-0.53}\times 10^{41}$   & 19/31 & $1.7^{+0.65}_{-0.58}$  \\
Peak        & $92^{+8.2}_{-7.4}$    &   $4.4^{+0.02}_{-0.68}\times 10^{41}$   & 42/31 & $2.2^{+0.20}_{-0.24}$  \\
Decay       & $71^{+10}_{-9.4}$     &   $2.8^{+1.3}_{-0.83}\times 10^{41}$   & 42/31 & $2.9^{+1.0}_{-1.1}$ \\
\end{tabular}
\\
\vspace{1em}
\textit{XMM} Jul-Aug 2020 (from \citet{Arcodia21})
\begin{tabular}{c|c|c|c}
\hline
Phase & $kT$ & $L_{bol}$ &  $R_{bb}$\\
 & (eV) & (erg s$^{-1}$) &  ($10^{10}$ cm) \\
\toprule
Quiesc.  & $130^{+33}_{-27}$ & $1.2^{+0.41}_{-0.23}\times 10^{41}$ & $0.57^{+0.15}_{-0.14}$  \\
XMM1 peak & $262^{+7}_{-6}$ & $4.1^{+0.15}_{-0.10}\times 10^{43}$ & $2.6^{+0.52}_{-0.44}$  \\
XMM2 peak & $148^{+8}_{-5}$ & $1.3^{+0.05}_{-0.14}\times 10^{43}$  & $4.6^{+0.3}_{-0.2}$ 
\end{tabular}
\caption{Spectral fitting results from the \textit{XMM-Newton} DDT observation on Jan 5-6 2024 (above) and comparison to the previously reported results in \citet{Arcodia21}. Errors are $1\sigma$, and the model is \texttt{tbabs}$\times$\texttt{zbbody}. Compared to the previous \textit{XMM}-detected QPEs, the peak temperatures and luminosities are now significantly lower. On the other hand, due to the low count rate (hence large errors) of quiescence, the quiescence $L_{bol}$ and $R_{bb}$ are within $3\sigma$ consistent with the previous \textit{XMM} detection in July-August 2020.}
\label{tab:xmm}
}
\end{center}
\end{table}

\subsection{Flare profiles}
\label{subsec:flare_profiles}

\nicer\ has observed eRO-QPE1 for several observing epochs lasting 6-15 days over the past 3.5 years. QPEs are visible in all epochs except Oct. 2023 (but have since emerged in Nov 2023), with complex, non-monotonically varying properties including QPE luminosity, recurrence time, duration, temperature, and the intra-epoch scatter of these properties. To allow for precise characterization of these properties, we fit each observed QPE with the following exponential flare model motivated initially for long gamma-ray bursts \citep{Norris05} and adopted for eRO-QPE1 in \citet{Arcodia22}:
\[ \mathrm{QPE\;flux} = \begin{cases} 
      A\lambda e^{{\tau_1/(t_{peak}-t_{as}-t)}} & \mathrm{if}\; t<t_{peak} \\
      Ae^{{-(t-t_{peak})/\tau_2}} & \mathrm{if}\; t\geq t_{peak}
   \end{cases}
\]
where $A$ is the flare amplitude; $t_{peak}$ is the time of peak flux; $\tau_1$, $\tau_2$ are the e-folding times of rise and decay, respectively; $\lambda=e^{\sqrt{\tau_1/\tau_2}}$ is a normalization to join rise and decay; and $t_{as}=\sqrt{\tau_1 \tau_2}$ sets the asymptote time such that flux$=0$ for $t<t_{peak}-t_{as}$. We compute integrated energy outputs of the flares \textcolor{black}{(Section~\ref{section:secular_evolution})} by integrating from ($t_{peak}-3\tau_1$) to ($t_{peak}+3\tau_2$), i.e. the total energy contained within $\pm 3$ e-folds of the flare peak. The observed flares are generally well-described by this exponential profile (Fig.~\ref{fig:lc}), and we use these fits to produce most of the results below.

\citet{Arcodia22} found that one \xmm\ observation of eRO-QPE1 (OBSID 0861910201) displayed multiple overlapping bursts, making it the only QPE source showing this behavior. One \nicer\ flare clearly shows this behavior (MJD 59413 in Jul 2021). The presence of at least two overlapping QPEs in eRO-QPE1, between \xmm\ 2020 and \nicer\ 2021, presents an interesting constraint for EMRI models, perhaps pointing towards a high-eccentricity EMRI orbit or unusual disk geometry. It is worth noting that the double-flare on MJD 59413 occurred unusually long after the previous burst a MJD 59412---a difference of 1.28 days, which is slightly under twice the mean recurrence time of nearby epochs. Thus, a geometry in which one burst was missed, then appeared alongside the next QPE, is possible. The more poorly-sampled QPE at MJD 59900 may be another instance of this double-flare behavior associated with a longer recurrence time, though that case is not as clear.

\subsection{Radio nondetections}

In addition to the \nicer\ campaign, we undertook radio monitoring of eRO--QPE1. Our campaign consisted of $2\times1{\rm \, hour}$ observations with MeerKAT in Sep 2021, and $7\times20 {\rm \, minute}$ observations with the Karl G. Jansky Very Large Array (VLA; see \S\ref{sec:vlarad} for details), partially concurrent with our Jan 2023 \nicer\ campaign. No observations resulted in a radio detection, though we note that unfortunately none of our snapshots were taken during an eruption (Fig.~\ref{fig:lc}). The $3\sigma$ upper limits are listed in Table~\ref{tab:radio}. Our most constraining radio limit comes from the stacked image of all seven VLA observations; $3-\sigma$ upper-limit of 15.6 $\mu$Jy at 6 GHz.

\subsubsection{MeerKAT observation details}
\label{sec:meerkatrad}

The MeerKAT radio telescope is a 64-dish interferometer based in the Karoo Desert, South Africa. We obtained two observations through a Director's Discretionary Time proposal (PI: R. Arcodia, DDT-20210908-RA-01). The observations were made on the 11\textsuperscript{th} and 19\textsuperscript{th} September 2021, each lasting 2.33\,hours. MeerKAT observes at a central frequency of 1.28\,GHz with a total bandwidth of 0.86\,GHz. For each observation, we started with five-minutes observing the flux and bandpass calibrator J0408-6545 followed by switching between the target for 40 minutes and the phase calibrator J0240-2309 for two minutes.

Both epochs were reduced using \textsc{oxkat}, a set of python scripts specifically designed for the reduction of MeerKAT observations \citep{Heywood20}. The scripts apply flags to the calibrator fields before calculating delay, bandpass and complex gain calibrations using both calibrator sources and applies them to the target, all of which is done in \textsc{casa}. The target field is split out and flagged then imaged using \textsc{wsclean}. For these observations, we also performed phase-only self-calibration. 

We did not detect any radio emission at the position of eRO-QPE1 in either epoch. We provide 3$\sigma$ upper limits in Table \ref{tab:radio}.

\subsubsection{VLA observation details}
\label{sec:vlarad}
To efficiently sample the X-ray QPE cycle, we ran simulations based on previous X-ray light curves of eRO--QPE1 to estimate the combination of observation number/separation that gave us the highest probability of a VLA observation occurring during an X-ray eruption. The result was 7 observations separated by 5--15 hours. Therefore, we observed eRO--QPE1 with the VLA (Project Code: 23A--059) for 7 epochs in 2023 January. 
The array was in the B configuration at the time of all of our observations. We used the 8-bit samplers, observing in C (4 -- 8 GHz) band, comprised of 2 base-bands, with 8 spectral windows of 64 2-MHz channels each, giving a total bandwidth of 1.024 GHz per base-band. We reduced and imaged the data within the Common Astronomy Software Application (\textsc{casa} version 5.6), using standard procedures outlined in the \textsc{casa}Guides\footnote{\url{https://casaguides.nrao.edu/index.php/Karl\_G.\_Jansky_VLA_Tutorials}.} for VLA data reduction (i.e., a priori flagging, setting the flux density scale, initial phase calibration, solving for antenna-based delays, bandpass calibration, gain calibration, scaling the amplitude gains, and final target flagging). For all observations, we used 3C147 (J0542+498) as a flux/bandpass calibrator, and J0239--0234 as a phase calibrator. Imaging was performed with natural weighting to maximize sensitivity.

\begin{center}
\begin{tabular}{c|c|c|c} 
Inst. & Frequency & MJD & $3\sigma$ UL ($\mu$Jy) \\
\hline\hline
MeerKAT & 1.28 GHz & 59468.15 & $<27$ \\
MeerKAT & 1.28 GHz & 59476.13 & $<30$ \\
VLA     & 6 GHz & 59963.06 & $<32.4$ \\
VLA     & 6 GHz & 59964.11 & $<27.9$ \\
VLA     & 6 GHz & 59967.07 & $<36.0$ \\
VLA     & 6 GHz & 59968.08 & $<87.9$ \\
VLA     & 6 GHz & 59971.90 & $<41.4$ \\
VLA     & 6 GHz & 59973.07 & $<33.0$ \\
VLA     & 6 GHz & 59974.04 & $<33.6$
\label{tab:radio}
\end{tabular}
\end{center}

\begin{figure*}[ht]
    \centering
    \includegraphics[width=\linewidth]{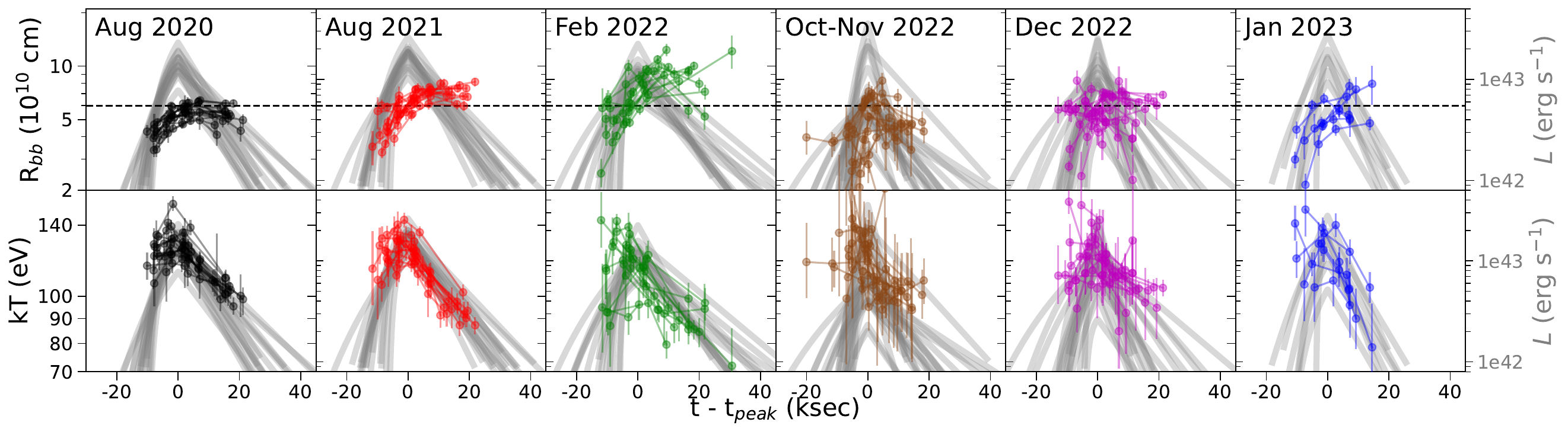}
    \caption{\textbf{Top:} Inferred blackbody radius $R_{bb}$ (filled points) plotted over \textcolor{black}{fitted double-exponential model} luminosity (gray), folded within each epoch. The short-term evolution is consistent with a cooling emission region expanding by a factor $\sim 2-3$ over each flare, in agreement with \citet{Miniutti23a}. Over the long-term, the peak inferred $R_{bb}$ appears to grow (from Aug 2020 to Feb 2022), then shrinks (by Oct-Nov 2022). This may be a geometric (viewing-angle) effect attributed to the EMRI apsidal precession ($\sim 10$s of days), or a change in the mutual EMRI-disk inclination due to the EMRI nodal precession ($\sim 100$s of days) resulting in changing ejecta properties (see discussion in Section~\ref{subsec:evolution_discussion}). \textbf{Bottom:} Same plot for $kT$.
     }
    \label{fig:rbb_kT}
\end{figure*}

\section{Results}
\label{sec:results}

\subsection{Short-term and secular evolution of the inferred emitting region}
\label{subsec:evolution_results}

QPEs show differing evolution at different energies, with narrower profiles and earlier peaks in higher-energy bands \citep{Miniutti19,Giustini20,Arcodia21,Chakraborty21,Quintin23,Arcodia24}. Said another way, \citet{Miniutti23a} found that QPEs in GSN 069 show $L-kT$ hysteresis, i.e. the blackbody temperature peaks early and begins to decay before the luminosity peaks. This is physically consistent with the inferred blackbody radius \textcolor{black}{(assuming a spherical source)}:
$$R_{bb}=\sqrt{\frac{L}{4\pi \sigma_{SB}T^4}}$$
expanding over time (shown in their Fig. 18). The QPEs in GSN 069 are consistent with an initial $R_{bb} \approx R_\odot$, expanding by a factor $\sim 3$ over the course of the flares \citep{Miniutti23a}. \citet{Arcodia22} found a similar hysteresis pattern is obeyed in the \xmm\ outbursts of eRO-QPE1 (their Figs. 7 \& 8). The longer temporal coverage of \nicer\ compared to \xmm\ allows us to track the average evolution of the inferred blackbody emission over the course of many QPEs, as well as how this quantity evolves in the long-term---an important constraint for distinguishing between theoretical models.

The early \nicer\ observations of eRO-QPE1 broadly agree with these trends, but this does not hold true at later epochs. In Aug 2020 and Aug 2021, the inferred blackbody radius $R_{bb}$ indeed appears to grow over the course of each QPE by a factor $\sim 2-3$ while the emission temperature cools by $\sim$50\% (Fig.~\ref{fig:rbb_kT}, also Figs.~\ref{fig:rbb_all} \& \ref{fig:kT_all}). Then, there is long-term evolution in the typical value of $R_{bb}$, which peaks around $6\times 10^{10}$ cm in Aug 2020 but increases up to $>10^{11}$ cm by Feb 2022, before returning to the previous value in subsequent epochs. The typical rate of change $dR_{bb}/dt$ correlates with the typical peak $R_{bb}$ in each epoch (though this cannot be directly interpreted as the physical expansion speed of the emission region; see Section~\ref{subsec:evolution_discussion}). The Aug 2020 and Aug 2021 epochs, which observed 15 QPEs each, show remarkable consistency in the evolution of $R_{bb}$ (as well as the flare profiles themselves), but this grows more erratic in later epochs, with larger short-term variation in typical QPE profiles, luminosities, $R_{bb}$, etc.

\begin{figure}
    \centering
    \includegraphics[width=\linewidth]{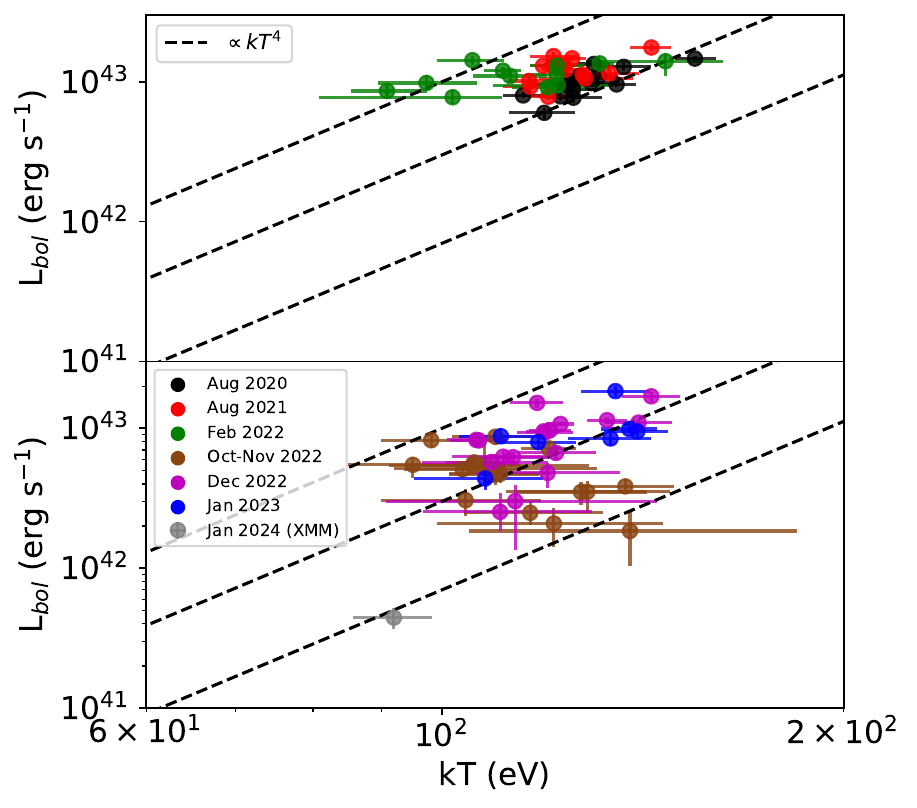}
    \caption{Peak $L_{bol}$ vs. $kT$ for each QPE. In general the trend appears consistent with $L_{bol} \propto T^4$, with the exception of the Feb 2022 and Oct-Dec 2022 epochs. The Oct-Dec 2022 epochs are also where $R_{bb}$ appears to invert its evolution (Fig.~\ref{fig:rbb_kT}). The $\propto T^4$ lines are plotted for visualization only, and are not fits to the data.}
    \label{fig:kT_Lbol}
\end{figure}

\subsection{Secular evolution of other QPE properties}
\label{section:secular_evolution}

\citet{Miniutti23b} noted the QPEs of GSN 069 roughly follow $L_{bol} \propto T^4$ at peak, which suggests that the X-rays comprise the majority of the emission of QPEs and that the peak $R_{bb}$ is roughly constant across flares. The peaks of eRO-QPE1 do not appear to obey this trend across all epochs (Fig.~\ref{fig:kT_Lbol}). At the lower luminosities typical of the Oct-Dec 2022 epochs, this relationship is less precisely followed, indicating the emission at lower $L_{bol}$ may not be consistent with pure blackbody, or a significant fraction is emitted below soft X-ray energies. Moreover, as clearly visible in Fig.~\ref{fig:rbb_kT}, the typical peak $R_{bb}$ is far less consistent in the Oct-Dec 2022 epochs, also contributing to the breaking of the $L\propto T^4$ trend.

\begin{figure}
    \centering
    \includegraphics[width=\linewidth]{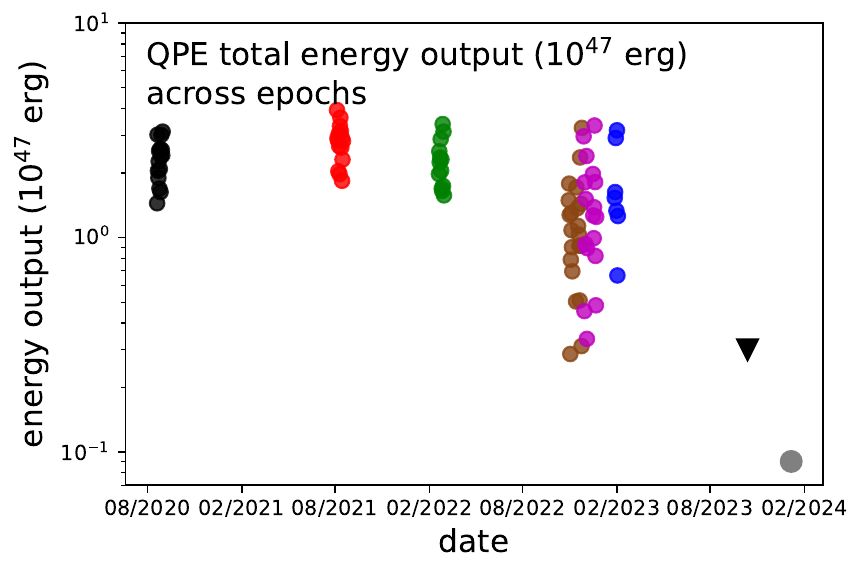}
    \caption{Secular evolution of QPE energy output (luminosity integrated over +/-3 e-folds from peak). The most recent epochs appear to show an overall decreasing trend, and QPEs are not seen in Oct 2023. We compute the Oct 2023 upper limit of the total energy by assuming the duration of possible QPEs stays roughly consistent with previous epochs. The peak of one QPE is seen in Nov 2023, but we cannot estimate the total energy output without better sampling of the overall flare.}
    \label{fig:energy}
\end{figure}

\begin{figure}
    \centering
    \includegraphics[width=\linewidth]{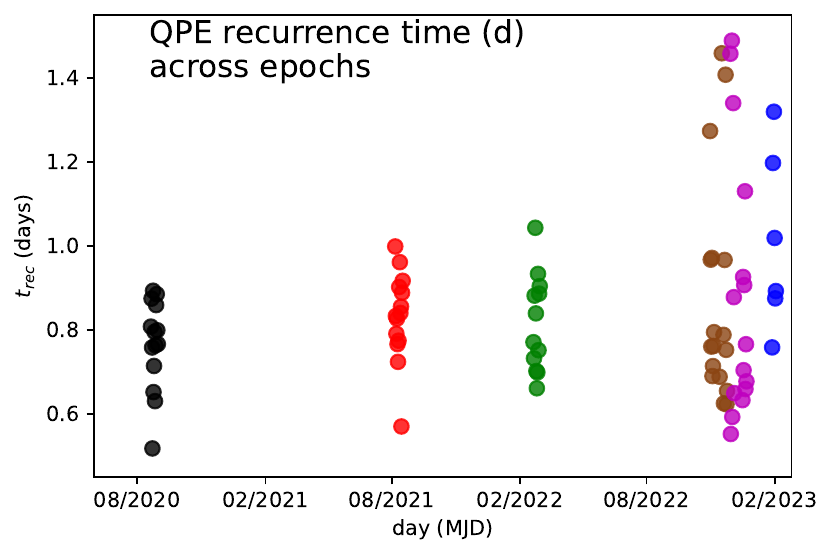}
    \caption{The QPE recurrence time $t_{rec}$ shows significant long-term evolution, varying between 0.76 and 1.10 days across 3 years (and doing so non-monotonically, possibly hinting at a longer-term precession period of $\sim 10-100$s of days).}
    \label{fig:t_rec}
\end{figure}

The increasingly erratic behavior of $R_{bb}$ (Fig.~\ref{fig:rbb_kT}) is also accompanied by an overall decrease in the typical QPE energy output (Fig.~\ref{fig:energy}), as well as growing irregularity in the flare recurrence times (Fig.~\ref{fig:t_rec}). The typical peak $R_{bb}$ increases by a factor of 3 over these years, before returning to its initial level; and the expansion speed $dR_{bb}/dt$ grows with the peak $R_{bb}$. We do not see any evolution in the typical decay times over the \textcolor{black}{3.5}-year baseline (Fig.~\ref{fig:decay_lpeak}), though the \nicer\ sampling cannot constrain the rise times precisely enough to make any comment about their change.

\textcolor{black}{The single QPE detected by \textit{XMM-Newton} in Jan 2024 (Fig.~\ref{fig:xmm}) appears at a flux $\sim$10x lower than the most recent \nicer-detected flares of 2023(Fig.~\ref{fig:kT_Lbol}), and $\sim$100x fainter/2.8x cooler than the previous \textit{XMM}-detected QPEs of July-August 2020 \citep{Arcodia21}. The most recent \textit{XMM} flare also exhibits an expanding emitting region, though its peak at $2.2\times 10^{10}$ cm (Fig.~\ref{fig:xmm}) is a factor of a few smaller than the typical $R_{bb}$ of the \nicer\ QPEs (Fig.~\ref{fig:rbb_kT}). Due to the large uncertainties of the quiescence spectra, the quiescence $L_{bol}$ and $R_{bb}$ are $3\sigma$ consistent with the previous \textit{XMM} detection in July-August 2020 (Table~\ref{tab:xmm}). This is in contrast to GSN 069, where a drop in the QPE luminosity was seen alongside a significant rise in the quiescence luminosity \citep{Miniutti23a}.}

\subsection{Evidence for a $\sim$6-day modulation in the QPE timing residuals}
\label{subsec:timing_results}

\subsubsection{QPE timing analysis methods}
\label{subsec:timing_methods}

After fitting the flares as described in Section~\ref{subsec:flare_profiles}, we use the computed $t_{peak}$ values to perform an O-C (Observed Minus Computed) analysis to search for any significant patterns in the timing behavior of the QPEs. The O-C technique, which is best known for its use in discovering the orbital period decay of the Hulse-Taylor binary pulsar \citep{Taylor82}, computes the timing deviation of a recurring signal with respect to a constant period. The residuals are computing by subtracting the expected (Computed) arrival time of the $N$th strictly periodic repetition (with period $P_{const}$) after some fixed starting time, from the Observed actual arrival time of that repetition. In the case of gravitational-wave driven orbital decay, these O-C values are fit with a quadratic because of $\dot{P}<0$; thus, the actual transit times arrive successively earlier and earlier than expected from naive extrapolation of $P_{const}$.

In our case, we separately compute an O-C within each epoch \textcolor{black}{because we cannot phase-connect across them---i.e., we have no way to tell how many QPEs  have occurred  between epochs, and the average recurrence time also changes significantly ($\sim 30$\%) between them (Fig.~\ref{fig:t_rec}) for reasons not yet physically understood}. Nevertheless, we are interested in probing the short-timescale behavior of the QPE timings independently of the secular evolution. Thus, on the timescale of individual \nicer\ epochs, we assume a $P_{const}$ given by the average recurrence time within that epoch, then compute O-C as described above.

As the O-C timing residuals are roughly sinusoidal (Sec.~\ref{subsec:oc_results}), we perform a least-squares fit to the residuals with a sine+constant model. We choose not to apply a Fourier-based technique to estimate the power spectral density (PSD) for two reasons: 1) our \nicer\ observing windows of 6-15 days limit us to only 1-2 periods per epoch, and 2) as the QPEs ``sample" the sinusoidal profile only on their recurrence time ($\sim$1 day), we have too few data points to make a meaningful estimate of the PSD. Comparing our sine+constant fit to a constant-only fit allows us to make a rough estimate of the recurrence period while remaining in the time domain. We fit each epoch individually, then all epochs together with a fixed overall period (but amplitude, phase, and constant offset free to vary within each epoch) to assess improvement over a constant model hypothesis. Our fit results are presented in \textcolor{black}{Section~\ref{subsec:oc_results}}, and the implications are discussed in Section~\ref{subsec:precession}.

\subsubsection{Results of O-C analysis}
\label{subsec:oc_results}

\begin{figure*}
    \centering
    \includegraphics[width=\linewidth]{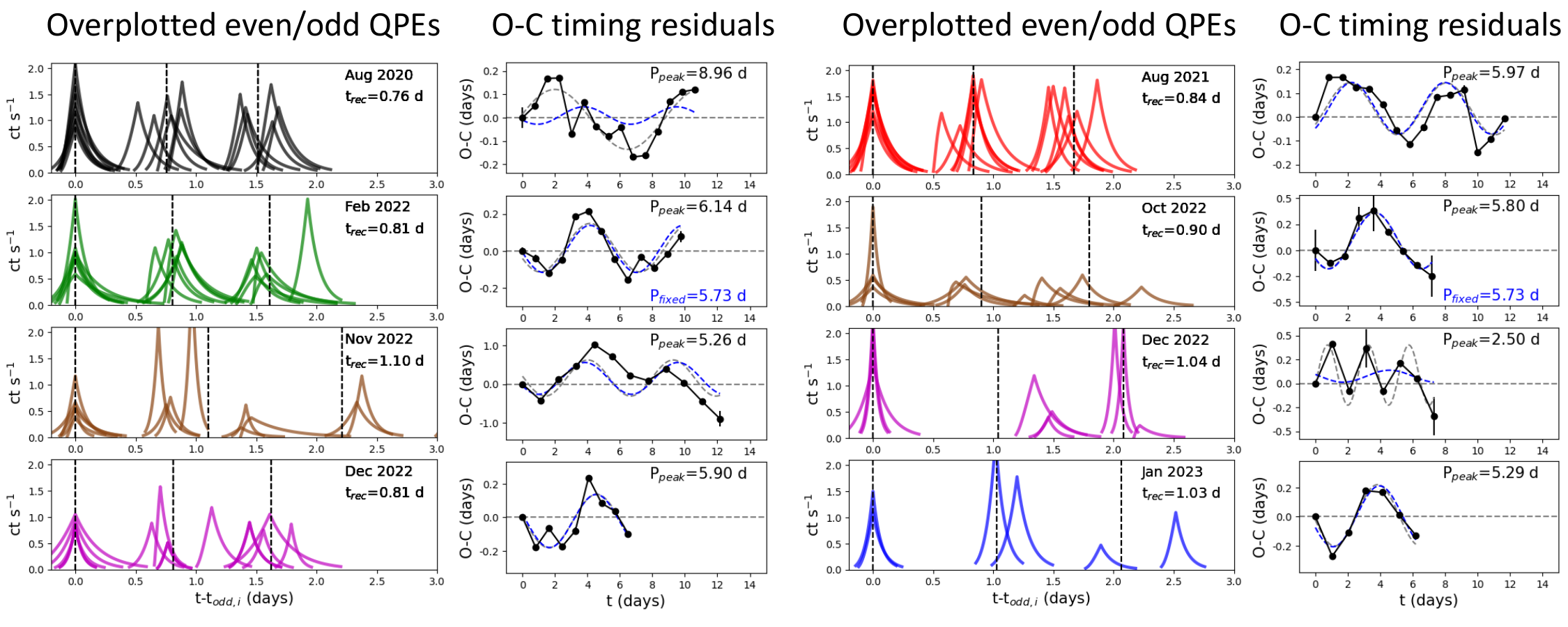}
    \caption{Overplotted even/odd burst sequences (demonstrating the high scatter and lack of any long/short recurrence pattern), and the corresponding O-C timing residuals (i.e. whether each burst arrives early/late compared to the average $t_{rec}$ within the epoch, Section~\ref{subsec:timing_methods}). As noted in \citet{Arcodia22}, the scatter in recurrence time ($\sim 50$\%) is significantly higher than other QPE sources. The O-C plots are overplotted with best-fit periods per-epoch (gray dashed line) and overall (blue dashed). The Aug 2020 epoch ($P_{peak}=8.96$ days) and first Dec 2022 epoch ($P_{peak}=2.5$ days) are significant outliers from the 5.73 average period. \textcolor{black}{The long-short recurrence pattern seen in other QPEs is not generally seen in eRO-QPE1, with the \textcolor{black}{possible} exception of a short-lived phase in early Dec 2022}.}
    \label{fig:timing}
\end{figure*}

The bursts of eRO-QPE1 show significant scatter compared to other QPEs \citep{Arcodia22}. To assess the degree of irregularity in the \nicer\ data, we folded the QPEs within each epoch into segments of three consecutive bursts, then overplotted each of these segments beginning at the time of the first burst (Fig.~\ref{fig:timing}). The QPE timings deviate by $\sim 50$\% from the average recurrence time computed in each epoch (compared to $\sim 10$\% in GSN 069 and eRO-QPE2).

The recurrence times do not correlate with any other properties of the flare (luminosity, rise time, and decay time), nor do those properties strongly correlate with each other (see Appendix~\ref{appendix:correlations}). Moreover, the average recurrence time varies across epochs, appearing to increase slightly from 0.76 days in Aug 2020 to 1.03 days in Jan 2023 (Fig.~\ref{fig:t_rec}).

We also report a tentative finding in the timing structure of the bursts (based on the procedure described in Section ~\ref{subsec:timing_methods}): the timing residuals (i.e. bursts arriving early/late compared to a strict period) exhibit possible periodic structure within several \nicer\ epochs, at an average period of 5.73 days (Fig.~\ref{fig:timing}). The best-fit period of the timing residuals are fairly consistent for most epochs (ranging from 5.2-6.1 days). Our fits of a sine+constant model provide a significant improvement over a constant model hypothesis, with an average $\Delta\chi^2=935$ in the per-epoch fits (gray dashed lines in Fig.~\ref{fig:timing}) and a total $\Delta\chi^2=5881$ in the global fixed-period fit (blue dashed lines).

The Aug 2020 and first Dec 2022 epochs are significant outliers, with best-fit super-periods of 9 days and 2.5 days respectively. Thus we cannot definitively claim a $\approx 6$ day super-period, but it is noteworthy, and follow-up observations will provide more clarity. Interestingly, a super-orbital period of $\sim$6 days has been predicted in some EMRI precession models, which we will describe in the discussion Section~\ref{subsec:precession}.

\section{Discussion}
\label{sec:discussion}

Our key observational findings in Section~\ref{sec:results} are:
\begin{itemize}
    \item The QPEs show secular evolution in $R_{bb}$, total energy output, and recurrence times (Figs.~\ref{fig:rbb_kT},~\ref{fig:energy},~\ref{fig:t_rec}) over the epochs from Aug 2020-Jan 2023. As seen in other sources, the QPEs are consistent with an emission region that grows in size and cools in temperature over the course of a flare. The characteristic peak $R_{bb}$ changes by a factor $\sim$2 over a timescale of years, and the emitting area evolution becomes increasingly erratic.
    \item QPEs are entirely undetected by \nicer\ in the Oct 2023 epoch, but the peak of one flare is seen in Nov 2023 (Fig.~\ref{fig:lc}). \textcolor{black}{We confirmed the ongoing presence of QPEs with \textit{XMM-Newton} in Jan 2024 (Fig.~\ref{fig:xmm}) at a luminosity $\sim 10$x lower than recent \textit{NICER} detections (Fig.~\ref{fig:kT_Lbol}), and $\sim$100x lower than the initial discovery \citep{Arcodia21}}
    \item The QPE recurrence times in most epochs are modulated on a $\sim$6-day super-period, and \textcolor{black}{the long-short recurrence pattern of other QPEs is generally not seen with the exception of a brief window in early Dec 2022} (Fig.~\ref{fig:timing}).
\end{itemize}

\subsection{QPE models}
\label{subsec:models}
There are a large number of models proposed to explain the origin of QPEs, fitting broadly into two categories. The first class is recurring limit-cycle instabilities within the SMBH accretion disk \citep{Raj21,Pan22,Pan23,Kaur23,Sniegowska23}, inspired by observations of recurring high-amplitude outbursts (the ``heartbeat" states) seen in black hole X-ray binaries GRS 1915+105 \citep{Belloni00} and IGR J17091-3624 \citep{Altamirano11}. In fact, Wang et al. 2024 reported high-amplitude variability at 0.5 Hz, which scaled up for a $10^5$ $M_\odot$ SMBH corresponds to $\sim 2$ days, similar to the QPE timescales. However, the largest weakness of these models is they make fewer testable predictions due to our poorer understanding of accretion instabilities in the radiation-pressure dominated regime.

The second class is interaction of the SMBH with a lower-mass orbiting companion in an extreme mass-ratio inspiral (EMRI), whether by direct accretion/tidal stripping of the companion \citep{King22,Metzger22,Krolik22,Zhao22,Linial23a,Lu23} or its interaction/collision with the SMBH accretion disk \citep{Xian21,Sukova21,Franchini23,Tagawa23,Linial23b}. EMRI-disk collision models have thus far been most successful at reproducing QPE observational properties within GSN 069 and eRO-QPE2, the two comparatively ``well-behaved" QPE sources. \textcolor{black}{Moreover, recent iterations of EMRI-disk models \citep{Franchini23,Linial23a} suggest the underlying accretion disk is comprised of debris from the disrupted star following a TDE---thus explaining the suggestive coincidence of QPEs with TDE hosts \citep{Chakraborty21,Miniutti23a,Quintin23}, as well as the compact emission radii and hot blackbody temperatures observed \citep{Miniutti23b}}. Thus, in the following discussion, we discuss our results in light of these EMRI-disk collision models.

GSN 069 and eRO-QPE2 obey a quasi-period to within 10\% \citep{Miniutti19,Arcodia21}, and also a characteristic ``long-short" recurrence time \citep{Miniutti23a}, which provided motivation for models to consider an EMRI companion on a mildly eccentric orbit \citep{Franchini23,Linial23b}. Within this picture, the alternating trend can be conveniently produced by associating the long recurrence times with the orbiter's passage through apocenter, and the short recurrence time through pericenter. Each EMRI-disk collision then ejects part of the disk mass, which is shock-heated by the supersonically moving EMRI. The disk ejecta, which is radiation-pressure dominated (as expected for the typical $\dot{M}$ of the quiescent disk), expands adiabatically over the QPE timescale, resulting in the observed declining $kT$ and increasing $R_{bb}$ (Fig.~\ref{fig:rbb_kT}).

eRO-QPE1, on the other hand, has proven difficult for these models to explain. In previous studies \citep{Arcodia21,Arcodia22}, this source has shown 1) significantly larger, apparently unpredictable scatter in burst recurrence times (up to 50\%, Fig.~\ref{fig:timing}); 2) apparently no long-short recurrence pattern; and 3) occasional overlapping double-flares (Section~\ref{subsec:flare_profiles}), raising concerns for the simple picture of a mildly eccentric EMRI generating QPEs twice per orbit. We may consider that the above models imply the QPE timing should be affected by several precession frequencies (as also discussed in \citet{Linial23b} and \citet{Franchini23}): \textcolor{black}{the underlying accretion disk} should precess nodally via gravitomagnetic frame dragging \citep{Lense18,Kato90,Merloni99}, and \textcolor{black}{the EMRI companion itself} should also precess apsidally ($P\sim$tens of days) and nodally ($P\sim$hundreds of days) \citep{Franchini23,Linial23b}. The precise observational implications of these precession effects has not yet been explored extensively, and we do so here.

\subsection{Secular evolution}
\label{subsec:evolution_discussion}

The long-term evolution in the flare properties $L$, $kT$, $R_{bb}$, and $\frac{dR_{bb}}{dt}$ (Figs.~\ref{fig:rbb_kT}, ~\ref{fig:kT_Lbol}, ~\ref{fig:energy}) can be explored within the framework of EMRI-disk interaction models, where the QPE emission is generated by shocked disk material from collisions with the companion and the disk.

The amount of ejected disk material $M_{sh}$ is set by the interaction cross-section of the companion and disk, which should depend on the geometric cross-section of the companion if it is a star \citep{Sukova21,Linial23b} or the Bondi-Hoyle radius if the companion is a compact object \citep{Franchini23}. There should also be a dependence on the relative inclination of the companion and the disk (in the face-on impact scenario, the orbiter collides with the minimum amount of disk mass, whereas for more highly inclined impacts there is a longer impact duration, thus more ejected mass). The resulting ejecta will have a kinetic energy $\propto \vec{v}_{sh}^2$, where $\vec{v}_{sh}$ is the relative velocity between the orbiter and the disk. $\vec{v}_{sh}$ also depends on the relative EMRI-disk inclination: for face-on impacts, the (Keplerian) orbital velocity of the disk material is entirely perpendicular to the EMRI velocity, but in general the vector difference $\vec{v}_{sh}= \vec{v}_{disk} - \vec{v}_{EMRI}$ will have an inclination-dependent magnitude. The ejecta then expands adiabatically (driven by radiation pressure from the significant energy within the shock-heated disk material), cooling over time as its optical depth declines; this expansion of the shocked ejecta causes the observed increase in $R_{bb}$ (Fig.~\ref{fig:rbb_kT}, \citet{Miniutti23a}).

It is still an open question whether the emission itself is dominated by blackbody emission, bremsstrahlung, Compton up-scattering, or some other radiative process. \citet{Linial23b} noted that the rapid timescale of QPEs may be too short for the ejecta to reach thermal equilibrium, with photon production instead dominated by free-free emission. It is difficult to clearly distinguish the two with only X-ray observations, but deep longer-wavelength observations may eventually be able to identify the dominant mechanism. In either case, fitting a blackbody model to the X-ray data allows us to infer a rough scale of the physical emitting region (which, in the bremsstrahlung case, \textit{does not} correspond to the actual physical extent of the photosphere).

From Fig.~\ref{fig:rbb_kT} we observe that the typical $R_{bb}$, as well as expansion rate $\frac{dR_{bb}}{dt}$, appears to increase over the epochs, which appears to suggest an increasing $\vec{v}_{sh}$ (thus a changing relative inclination). We cannot exactly use this to constrain the relative EMRI/disk inclination angles, as there are a number of different configurations (with a prograde/retrograde disk/EMRI) that create the same trend. One natural way to explain a change in the relative velocity is by invoking nodal precession of the disk and nodal+apsidal precession of the companion. We discuss the observable effect of these timescales further in Section~\ref{subsec:precession}.

The most recent Jan 2024 \textit{XMM-Newton}-detected QPE is a factor $\sim$100x fainter than initial discovery \citep{Arcodia21}. Moreover, the peak temperature has decreased by a factor $\sim$3x (Table~\ref{tab:xmm}). It is worth noting, if this trend continues, QPEs may continue while ceasing to be detectable at X-ray wavelengths, instead appearing as an ultraviolet transient \citep{Linial23c}.

\subsection{Order in the recurrence times}
\label{subsec:precession}

One natural explanation for the scatter in recurrence times within the EMRI-disk interaction picture is (independent) precession of the EMRI and the disk. The GR effects at the hours-long periods of QPEs should significantly modulate the EMRI and disk orbits, via nodal precession (induced by gravitomagnetic frame dragging; \citet{Lense18,Kato90,Merloni99}) and apsidal precession on observable timescales. We first discuss the expected disk nodal precession period for a given set of system parameters, following the arguments of \citet{Franchini16} and \citet{Franchini23}. Assuming the angular momentum of the disk to be misaligned with respect to the SMBH spin axis (which is generally expected for TDE disks, for an isotropic distribution of disrupted stars), frame dragging causes differential precession of the disc radii. If the disc viscosity is smaller than the disc thickness, the propagation of these disturbances occurs in the so-called `bending waves' regime, allowing the disk to rigidly precess \citep{Franchini16} on a timescale that depends on the SMBH mass ($M_{BH}$) and spin ($a$), as well as the disk outer radius ($R_{\mathrm{out}}$) and surface density profile.

Assuming the disk has a power-law surface density profile, with index $p$ and normalization $\Sigma_0$:
$$\Sigma(R)=\Sigma_0(R/R_g)^{-p}$$
$$\Sigma_0 = \frac{M_d(2-p)}{2\pi R_g^2}\bigg[\bigg(\frac{R_{\mathrm{out}}}{R_g}\bigg)^{2-p} - \bigg(\frac{R_{\mathrm{ISCO}}}{R_g}\bigg)^{2-p}\bigg]$$

and the nodal precession frequency at a given orbital radius is given by:
$$\Omega_{LT}(r) = \frac{c^3}{2GM_{BH}}\frac{4a\big(\frac{R}{R_g}\big)^{-3/2} - 3a^2\big(\frac{R}{R_g}\big)^{-2}}{\big(\frac{R}{R_g}\big)^{3/2}+a}$$

the total disk precession frequency is $\Omega_{LT}(R)$ integrated along the radial extent of the disk, weighted by the angular momentum $L(R)=\Sigma(R)V(R)R=\Sigma(R)\Omega_K(R)R^2$, where $\Omega_K=(GM_{BH}/R^3)^{1/2}$ is the Keplerian orbital frequency:
$$\Omega_p = \frac{\int_{R_{\mathrm{ISCO}}}^{R_{\mathrm{out}}} \Omega_{LT}(R) \times \Sigma(R)\Omega_K(R)R^2 \times 2\pi RdR}{\int_{R_{\mathrm{ISCO}}}^{R_{\mathrm{out}}} \Sigma(R)\Omega_K(R)R^2 \times 2\pi RdR}$$

\begin{figure}
    \centering
    \includegraphics[width=\linewidth]{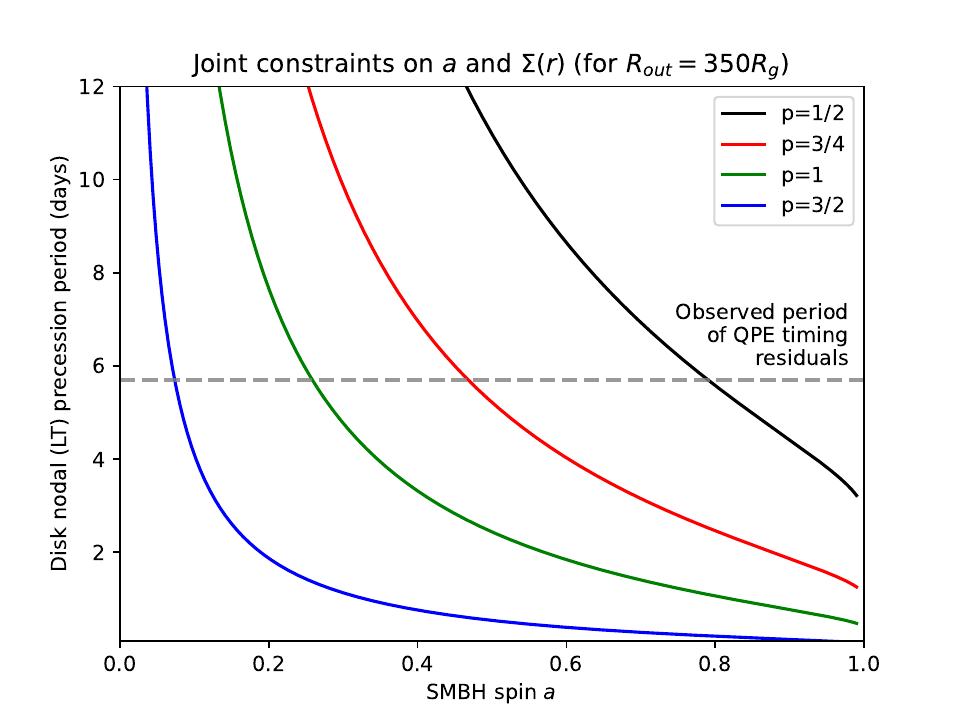}
    \caption{Joint constraints on SMBH spin $a$, and the power-law index $p$ of the disk radial density profile $\Sigma(R)\propto R^{-p}$. We assume $M_{BH} = 10^{5.8}$ M$_\odot$ \citep{Arcodia21}, and the accretion disk has outer radius $R_{\mathrm{out}}=350R_g$. These constraints depend strongly on the assumed $M_{BH}$ and $R_{\mathrm{out}}$.}
    \label{fig:a_p_constraint}
\end{figure}

Observationally, the effect of the disk nodal precession is to change the relative inclination between the disk and EMRI companion, thus delaying or advancing the impact/interaction times resulting in QPEs. If we assume our reported QPE timing residuals (Fig.~\ref{fig:timing}) are associated with the disk LT precession, for $M_{BH}\sim 10^{5.8}\;M_\odot$ \citep{Arcodia21}, $M_D=2M_\odot$, and $R_{\mathrm{out}}=100 R_g$, we can constrain (although degenerately) the SMBH spin $a$ and disk surface density profile $\Sigma(R)$ (Fig.~\ref{fig:a_p_constraint}). It is important to note that these constraints depend sensitively on $M_{BH}$ and $R_{\mathrm{out}}$, increasing either of which will increase the inferred disk precession period.

The timing residuals are not strictly sinusoidal (Fig.~\ref{fig:timing}), as expected due to the presence of other relevant frequencies which also modulate the burst arrival times. For example, three possible effects are light travel-time delays from apsidal precession of the EMRI ($P\sim$tens of days), changing relative EMRI-disk inclination due to nodal precession of the EMRI ($P\sim$hundreds of days) \citep{Linial23b,Franchini23}, or variable disk surface density due to stochastic mass accretion rate fluctuations as proposed for X-ray binaries \citep{Ingram11}. Still, the hierarchy of precession timescales is such that, over individual \nicer\ epochs ($\sim$10 days) the disk nodal precession should dominate, resulting in the observed $\sim$6-day period (largely consistent with the assumed 7.5-day nodal precession period used for the model of eRO-QPE1 in \citet{Franchini23}).

The apparent modulation of the QPE arrival times on a timescale consistent with the predicted disk precession period supports QPEs being generated by some direct interaction with the accretion disk \citep{Sukova21,Xian21,Linial23b,Franchini23,Tagawa23}. Other models that propose that QPEs are fed by the companion itself (i.e. where the quiescent disk does not play a part in modulating the timing of the burst) require some other tuning to account for the \textcolor{black}{erratic} modulation of the QPE period. The precise radiative processes generating the emission following the EMRI-disk interactions are still uncertain, and will require further observations to constrain.

Another observational consequence of disk precession is modulation of the quiescent emission (it is this reason, for example, that QPOs in accreting stellar-mass black holes are sometimes interpreted in light of rigid precession). However, since the quiescent level of eRO-QPE1 is below \nicer\ detectability, we are unable to verify this. The presence of a quiescent-level QPO in GSN 069, during epochs showing QPEs, is worth noting \citep{Miniutti23a}.

One possible shortcoming of the disk precession interpretation is the alignment timescale of the disk with the SMBH spin \citep{Franchini16}. As the disk cools (lower $\dot{M}$) and becomes thinner, it eventually violates the conditions necessary for warp propagation in the bending waves regime, therefore no longer allowing rigid precession. In addition, the intrinsic viscosity of the disk should also damp oscillations. The precession amplitude should thus decay over a timescale of $\sim$years, with the exact value dependent on the disk viscosity, SMBH spin, and initial misalignment angle. As these quantities are all fairly uncertain, \textcolor{black}{matching the observed} years-timescale precession decay would require some fine-tuning.  One interpretation of this is that the accretion disk in eRO-QPE1 is dynamically younger than those in e.g. GSN 069 or eRO-QPE2, where the recurrence pattern is comparatively well-behaved and shows the long/short recurrence of a mildly eccentric orbit, indicative of a relatively stable disk configuration. Indeed, the QPEs in GSN 069 appeared up to 8 years after the appearance of a TDE-like increase in quiescent flux \citep{Miniutti19}, suggesting the accretion system had time to evolve. No such constraints are available for eRO-QPE1, eRO-QPE2, \textcolor{black}{or eRO-QPE3,} which were first discovered during their QPE-exhibiting phases. \textcolor{black}{In eRO-QPE4, the quiescence must have brightened with or after the first-detected QPEs, inidicating a short-lived nature/evolution of the accretion flow \citep{Arcodia24}.}

The Aug 2020 and first Dec 2022 epochs are significant outliers from the $\sim$6-day super-period (Fig.~\ref{fig:timing}). The Aug 2020 timing residuals appear roughly sinusoidal, but with a significantly longer $P=8.96$ days. It is unclear what causes this deviation, though one possible explanation is a changing disk surface density profile (a shallower decay increases the LT precession period for a fixed SMBH spin, Fig.~\ref{fig:a_p_constraint}), or viscous spreading of the disk resulting in a longer precession period. The first Dec 2022 epoch shows a brief long-short recurrence phase, seen in other sources but \textcolor{black}{generally not in eRO-QPE1, possibly indicative of a lower eccentricity $e\approx 0$, but with large uncertainties considering other deviations due to precession effects or secular evolution}. It may be that this epoch occurred during a particular disk/EMRI configuration resulting in double-impacts, but was short-lived, thus returning to the $\sim$6-day period in successive Dec 2022 and Jan 2023 epochs. Still, we emphasize that these explanations are speculative, and the reason these epochs are outliers is very unclear. An alternative explanation to the disk/EMRI precession picture is the presence of a third body in the system, e.g. a second solar-mass companion in orbit around the SMBH which modulates the orbital frequency/configuration between epochs.

\section{Conclusions}
\label{sec:conclusions}

We analyzed data from the \textcolor{black}{3.5}-year \nicer\ campaign of eRO-QPE1, with the following key findings:
\begin{itemize}
    \item The QPEs show complex, non-monotonic secular evolution, including variations by a factor of several in the characteristic luminosities, temperatures, blackbody radii, total energy output, and recurrence times. In the Oct 2023 epoch, the QPEs disappeared within \nicer\ detectability (Fig.~\ref{fig:lc}), \textcolor{black}{but we confirmed they are ongoing with \textit{XMM-Newton} in Jan 2024 (Fig.~\ref{fig:xmm}), at a level $\sim$10x fainter than recent \textit{NICER} detections and $\sim$100x fainter (and $\sim$3x cooler) than initial discovery \citep{Arcodia21}}. The increasing irregularity of QPE recurrence times (Fig.~\ref{fig:t_rec}) and $R_{bb}$ evolution (Fig.~\ref{fig:rbb_kT}) coincides with the decreasing energy output (Fig.~\ref{fig:energy}). \textcolor{black}{The large dynamic range of QPEs even within a single source is an important constraint for theoretical models.}
    \item The QPEs show $L-kT$ hysteresis (Fig.~\ref{fig:rbb_kT}) as also seen in \citet{Arcodia22}, which is physically consistent with a slowly-cooling emitting region which grows by a factor $\sim 2-3$ over the course of a flare. We infer $R_{bb}\sim R_\odot$, which is similar to the size for GSN 069 (\citet{Miniutti23a} Fig. 18). The long-term increase, then decrease of peak $R_{bb}$ may point to a long-term super-period caused by e.g. by the apisdal/nodal precession cycles of an EMRI companion, but any such periodicity will take multiple years of observation to confirm.
    \item Radio campaigns with MeerKAT and VLA did not detect the source at any epochs (Tab.~\ref{tab:radio}, Fig.~\ref{fig:lc}). Though the VLA campaign was partly simultaneous with a \nicer\ observing epoch (Jan 2023), none of the radio observations aligned with a burst; thus we are unable to conclude whether the QPEs are definitely undetected at radio wavelengths. The lack of detected radio emission during quiescence rules out the presence of strong AGN activity, thus the presence of a long-lived accretion disk.
    \item There is a possible $\sim$6-day modulation in the timing residuals of the bursts (Fig.~\ref{fig:timing}) during 6 out of 8 epochs. Interestingly, this period aligns with the inferred accretion disk nodal precession period. We may explain the significant scatter in QPE recurrence times, which is an unusual feature of eRO-QPE1, by a periodically changing relative inclination between the EMRI companion and the quiescent disk altering the relative delay/advance of each EMRI-disk interaction. This also provides a convenient explanation for the \textcolor{black}{general lack} of a long-short recurrence pattern (though it appears briefly in the first Dec 2022 epoch alongside a high-low amplitude, Fig.~\ref{fig:timing}), and the atypically long quasi-period of eRO-QPE1.
    \item Associating the $\sim$6-day modulation with the nodal precession frequency of the disk (due to gravitomagnetic frame dragging) allows us to make a (degenerate) constraint on the SMBH spin and disk radial density profile (Fig.~\ref{fig:a_p_constraint}). The result is sensitive to the choice of $M_{BH}$ and disk $R_{\mathrm{out}}$, and only weakly dependent on $M_D$.
\end{itemize}

The decreasing luminosity/altogether disappearance of QPEs (Fig.~\ref{fig:lc}) has been noted in other sources already \citep{Chakraborty21,Miniutti23a,Arcodia24}. Notably, in GSN 069 the QPEs re-appeared after one year \citep{Miniutti23b}, revealing a quiescent luminosity threshold for the appearance of QPEs and a dependence of the burst amplitude/temperature on the quiescent disk properties (tending asymptotically towards $kT_{QPE}/kT_{disk} = 1$ as the quiescent luminosity increases). \textcolor{black}{We cannot say definitively whether QPEs disappeared altogether in Oct 2023 (as \nicer\ is not sensitive enough to detect the quiescence), and indeed the presence of one peak in the Nov 2023 observations suggests they may have been ongoing at an undetectable faintness all along. The detection of a faint QPE in Jan 2024 with \textit{XMM-Newton} seems to agree with this, though the lack of a significant change in quiescence luminosity alongside the faint QPE (in contrast to GSN 069) now presents an open puzzle, complicating the dependence on the disk $\dot{M}$ obeyed by QPEs in general. Continued monitoring with \xmm\ and \nicer\ will reveal the evolution of eruptions in this eRO-QPE1, and whether more unexpected behaviors will continue to arise.}

\section*{Acknowledgements}
\textcolor{black}{We thank the anonymous reviewer for useful comments which improved the manuscript.} We thank Herman Marshall and Brian Metzger for useful discussions. R.A. received support for this work by NASA through the NASA Einstein Fellowship grant No HF2-51499 awarded by the Space Telescope Science Institute, which is operated by the Association of Universities for Research in Astronomy, Inc., for NASA, under contract NAS5-26555. G.M. was supported by grant PID2020-115325GB-C31 funded by MCIN/AEI/ 10.13039/501100011033. M.G. is supported by the ``Programa de Atracci\'on de Talento'' of the Comunidad de Madrid, grant number 2022-5A/TIC-24235. A.J.T. acknowledges partial support for this work provided by NASA through the NASA Hubble Fellowship grant \#HST--HF2--51494.001 awarded by the Space Telescope Science Institute, which is operated by the Association of Universities for Research in Astronomy, Inc., for NASA, under contract NAS5--26555. This work was supported by the Australian government through the Australian Research Council’s Discovery Projects funding scheme (DP200102471). \textcolor{black}{G.P. acknowledges financial support from the European Research Council (ERC) under the European Union’s Horizon 2020 research and innovation program HotMilk (grant agreement No. 865637), the Bando per il Finanziamento della Ricerca Fondamentale 2022 dell’Istituto Nazionale di Astrofisica (INAF): GO Large program, and the Framework per l’Attrazione e il Rafforzamento delle Eccellenze (FARE) per la ricerca in Italia (R20L5S39T9).} 

\appendix
\renewcommand\thefigure{\thesection.\arabic{figure}}    
\setcounter{figure}{0}

\section{\textit{NICER} source detection}
\label{appendix:detection}

As discussed in Section~\ref{subsec:nicer}, reliably estimating light curves for faint sources like eRO-QPE1, in which the source count rate is often comparable to the background, presents a significant challenge for \nicer. In several epochs, background variability can dominate over the source, thus making unclear whether a higher count rate is due to QPEs or the background. However, we have a strong prior set by our knowledge of the intrinsic source spectrum \citep{Arcodia21}, which we can exploit to identify the intrinsic source variability.

\begin{figure}[ht]
    \centering
    \includegraphics[width=\linewidth]{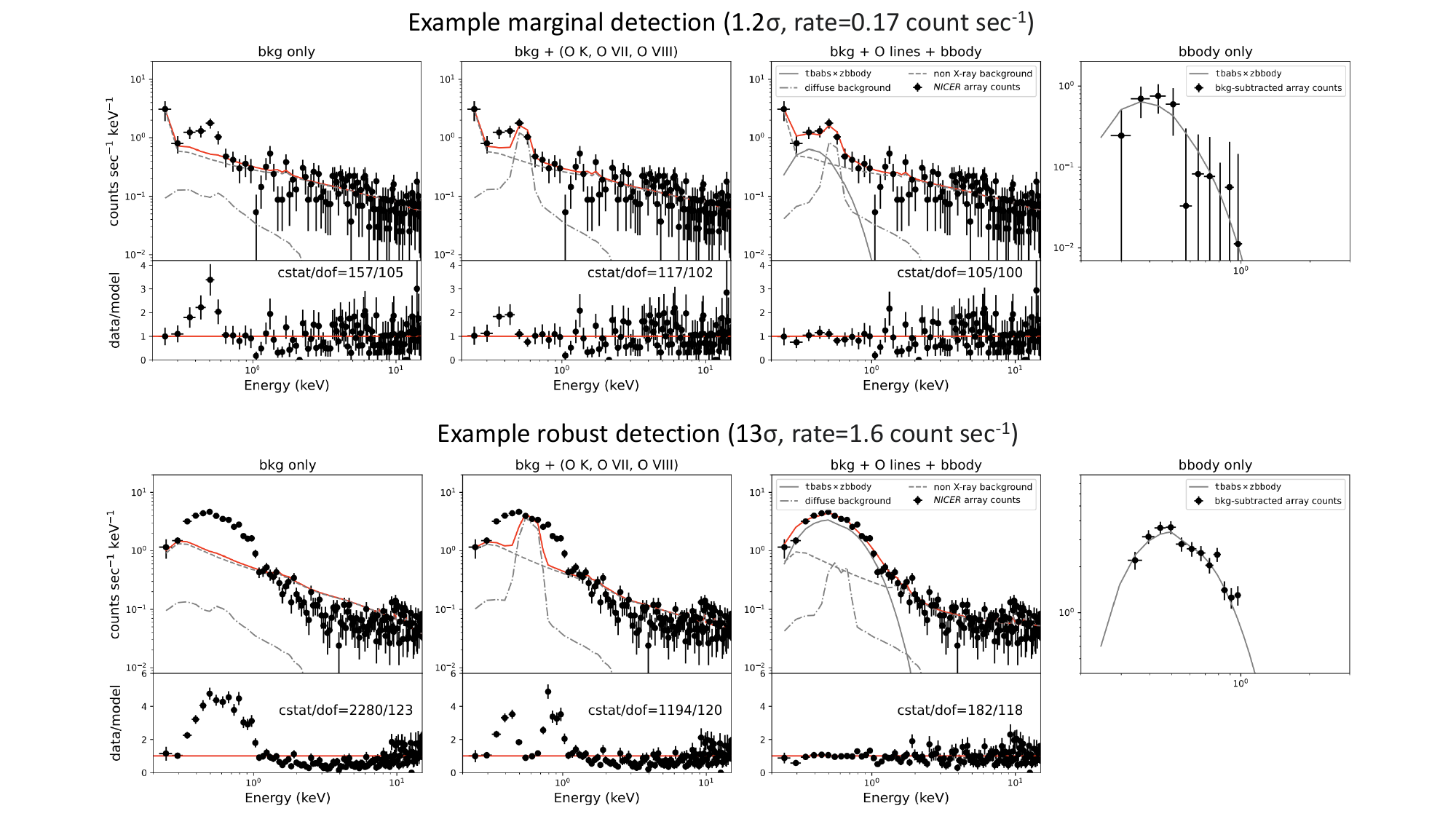}
    \caption{\nicer\ spectra for a sample marginal detection (1.2$\sigma$, top) and robust detection (13$\sigma$, bottom). We show spectra fit first with only the default \texttt{SCORPEON} components (representing the diffuse X-ray background and non X-ray background), then with the addition of ionized oxygen lines from the solar wind charge exchange (SWCX), then with the addition of a \texttt{tbabs$\times$zbbody} representing the source contribution.}
    \label{fig:detections}
\end{figure}

In Fig.~\ref{fig:detections} we show example GTIs for which the source was marginally or robustly detected. We consider as ``detections" any GTI in which a blackbody component is required, with normalization $>1\sigma$ inconsistent with zero. We used PyXspec to automate spectral fitting across a total of $\sim$5500 GTIs of $\leq$200 seconds, then summed the source-only counts in the 0.3-1 keV band (e.g. Fig.~\ref{fig:detections} right), to make the light curves shown in Fig.~\ref{fig:lc}.

\clearpage

\section{Complete radius and temperature evolution}
\label{appendix:rbb_kT_all}

\begin{figure}[ht]
    \centering
    \includegraphics[width=\linewidth]{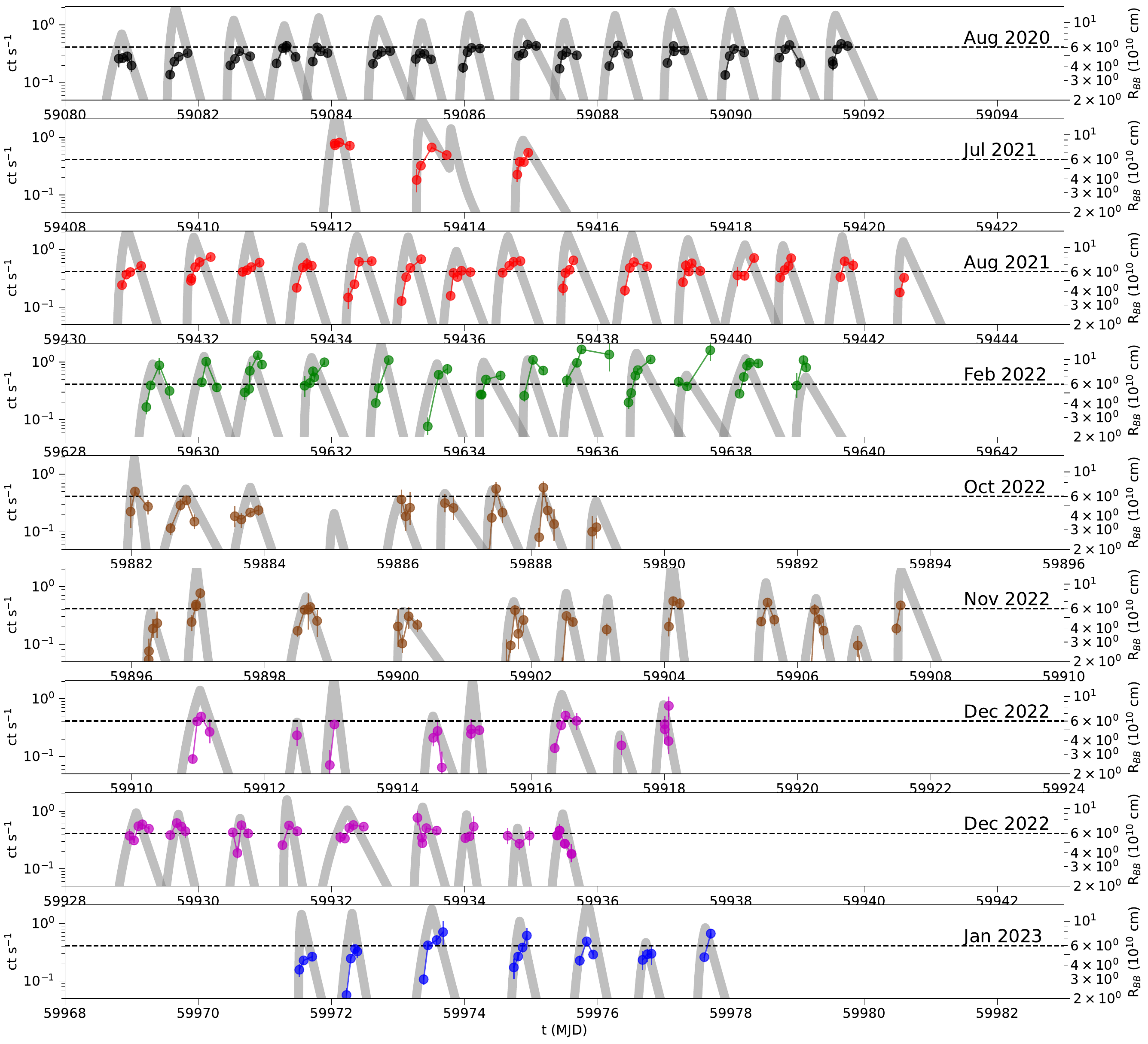}
    \caption{Evolution of blackbody radius ($R_{bb}$) for all QPEs, plotted separately.}
    \label{fig:rbb_all}
\end{figure}

\begin{figure}
    \centering
    \includegraphics[width=\linewidth]{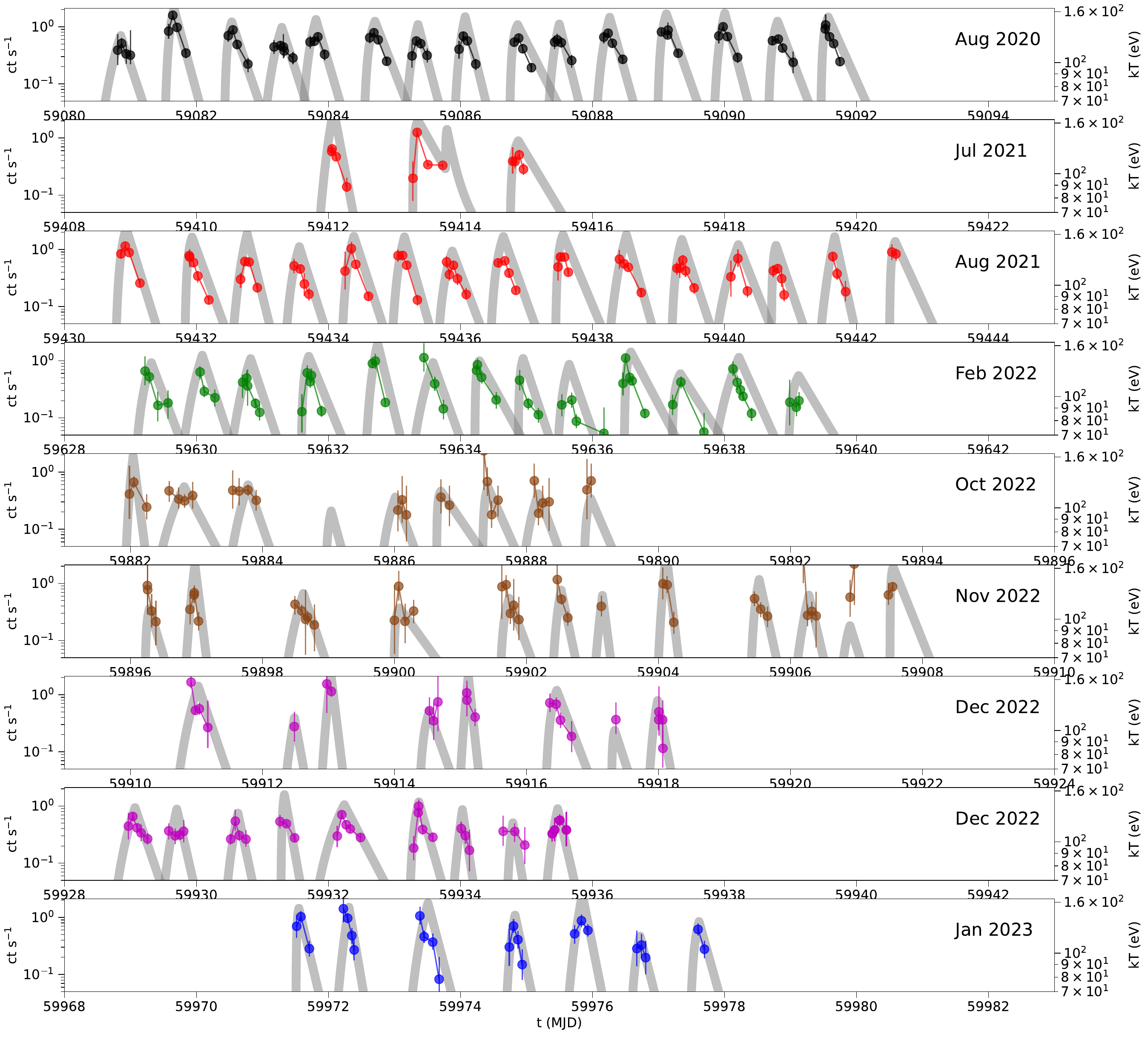}
    \caption{Evolution of blackbody temperatre ($kT$) for all QPEs, plotted separately.}
    \label{fig:kT_all}
\end{figure}

For completeness we present unfolded versions of Fig.~\ref{fig:rbb_kT}, showing the evolution of blackbody radius (Fig.~\ref{fig:rbb_all}) and temperature (Fig.~\ref{fig:kT_all}) for each individual QPE.

\clearpage

\section{QPE sample-wide correlations and energy distributions}
\label{appendix:correlations}

\begin{figure}[ht]
    \centering
    \includegraphics[width=\linewidth]{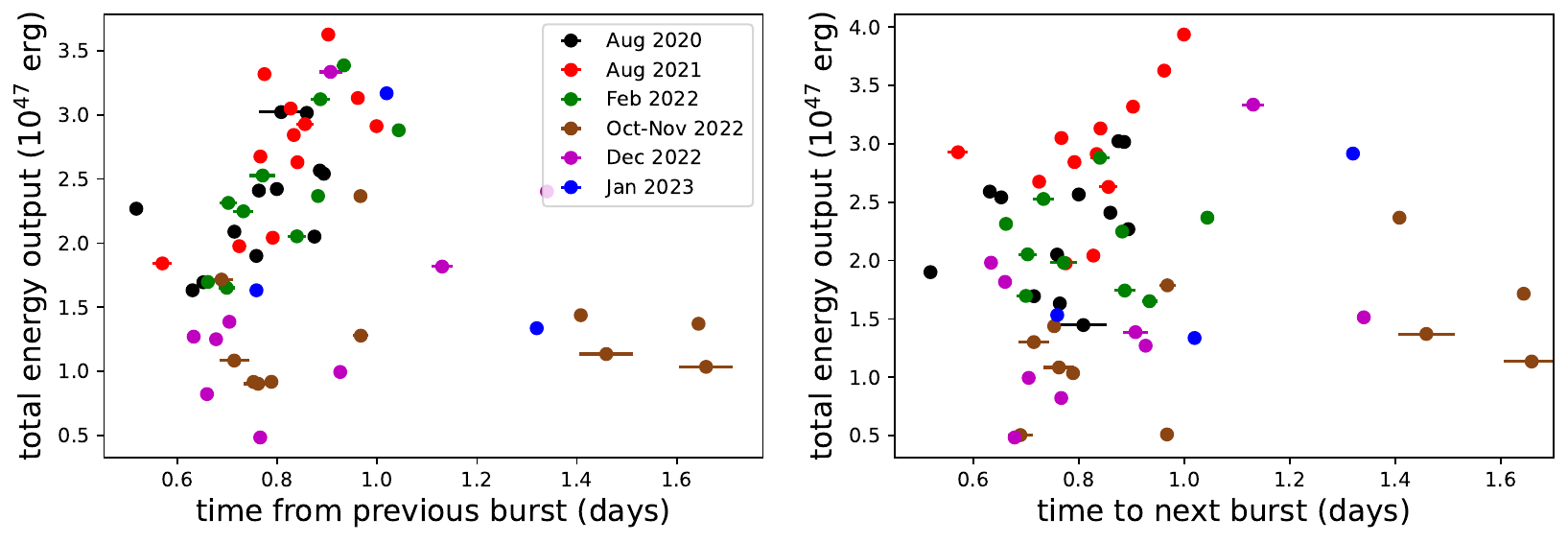}
    \caption{Total energy output of each QPE (y-axis) vs. time delay from previous burst (x-axis, left) or to next burst (x-axis, right). No clear correlation is apparent.}
    \label{fig:energy_trec}
\end{figure}

\begin{figure}
    \centering
    \includegraphics[width=0.6\linewidth]{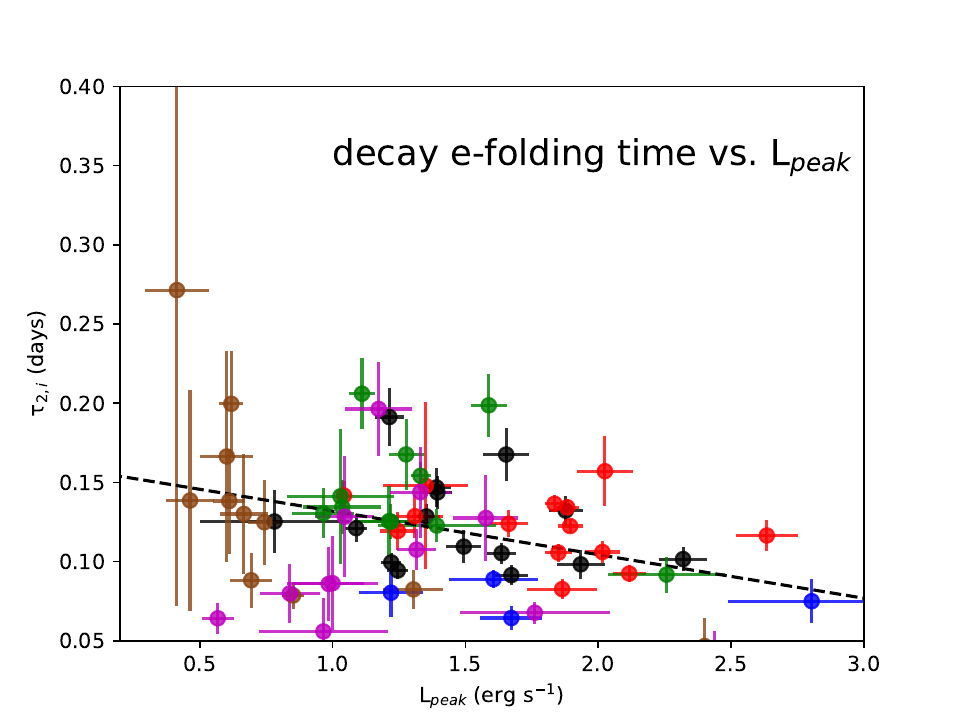}
    \caption{Decay e-folding time vs. peak QPE luminosity. There appears to be a slight anticorrelation with $L_{peak}$ and the decay time. The rise times are not plooted, as they are generally poorly constrained due to the sparser sampling, thus we cannot make strong statements about the presence/lack of a correlation. Where the rise times are constrained, they are typically a factor of $\sim 2$ shorter than the decay times.}
    \label{fig:decay_lpeak}
\end{figure}

\begin{figure}
    \centering
    \includegraphics[width=\linewidth]{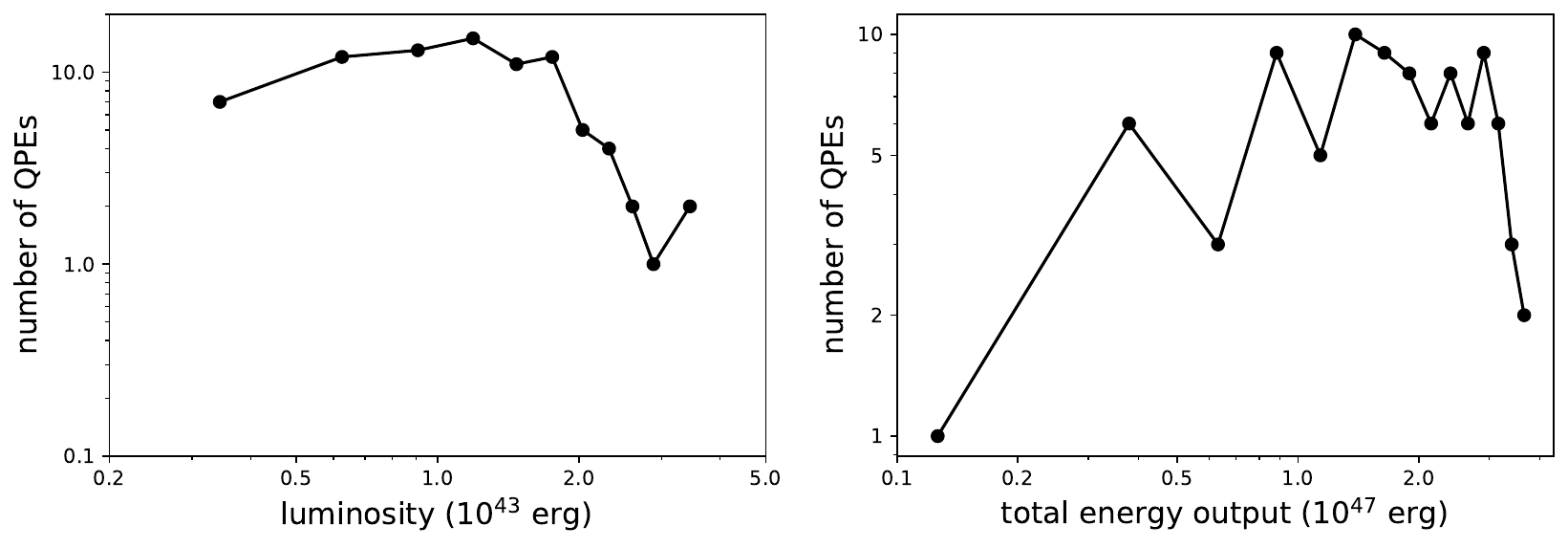}
    \caption{Distributions of QPEs by peak luminosity (left) and total integrated energy output (right).}
    \label{fig:lum_function}
\end{figure}

Here we present plots correlating the flare features across the population of QPEs across all epochs. Not all 92 QPEs are used in each plot, as some bursts are too poorly-sampled to determine e.g. rise/decay time, and thus total energy output. Fig.~\ref{fig:energy_trec} shows that there is no apparent correlation of the QPE total energy outputs with time before/after the burst. Fig.~\ref{fig:decay_lpeak} shows there is a slight anticorrelation of peak luminosity with decay time, though we cannot constrain a similar relationship for the rise times with the \nicer\ sampling. Fig.~\ref{fig:lum_function} shows the distributions of QPEs by peak luminosity and total energy output are roughly uniform.

\bibliography{refs}{}
\bibliographystyle{aasjournal}

\end{document}